# Detection of a Cooper-Pair Density Wave in $Bi_2Sr_2CaCu_2O_{8+x}$

M. H. Hamidian[1†], S. D. Edkins[2,3 †], Sang Hyun Joo[†4,5], A. Kostin[2], H. Eisaki[6], S. Uchida[6,7], M. J. Lawler[2,8], E.-A. Kim[2], A. P. Mackenzie[3,9], K. Fujita[10], Jinho Lee[4,5] and J.C. Séamus Davis[2,3,10,11]

1. Department of Physics, Harvard University, Cambridge, MA 02138, USA
2. LASSP, Department of Physics, Cornell University, Ithaca, NY 14853, USA
3. School of Physics and Astron., University of St. Andrews, Fife KY16 9SS, Scotland.
4. IAP, Department of Physics and Astron., Seoul National University, Seoul 151-747, Korea.
5. CCES, Institute of Basic Science, Seoul 151-747, Korea.
6. Inst. of Advanced Industrial Science and Tech., Tsukuba, Ibaraki 305-8568, Japan.
7. Department of Physics, University of Tokyo, Bunkyo, Tokyo 113-0011, Japan.
8. Department of Physics, Binghamton University, Binghamton, NY 13902-6000, USA
9. Max-Planck Institute for Chemical Physics of Solids, D-01187 Dresden, Germany.
10. CMPMS Department, Brookhaven National Laboratory, Upton, NY 11973, USA.
11. Kavli Institute at Cornell for Nanoscale Science, Cornell University, Ithaca, NY 14853, USA.

† These authors contributed equally to this project.

The quantum condensate of Cooper-pairs forming a superconductor was originally conceived to be translationally invariant. In theory, however, pairs can exist with finite momentum $Q$ and thereby generate states with spatially modulating Cooper-pair density[1,2]. While never observed directly in any superconductor, such a state has been created in ultra-cold $^6Li$ gas[3]. It is now widely hypothesized that the cuprate pseudogap phase[4] contains such a 'pair density wave' (PDW) state[5-21]. Here we use nanometer resolution scanned Josephson tunneling microscopy[22-24] (SJTM) to image Cooper-pair tunneling from a $d$-wave superconducting STM tip to the condensate of $Bi_2Sr_2CaCu_2O_{8+x}$. Condensate visualization capabilities are demonstrated directly using the Cooper-pair density variations surrounding Zn impurity atoms[25] and at the $Bi_2Sr_2CaCu_2O_{8+x}$ crystal-supermodulation[26]. Then, by using Fourier analysis of SJTM images, we discover the direct signature of a Cooper-pair density modulation at wavevectors $Q_P \approx (0.25, 0)2\pi/a_0; (0, 0.25)2\pi/a_0$ in $Bi_2Sr_2CaCu_2O_8$. The **amplitude of these modulations is ~ 5% of the homogeneous condensate density and their form factor exhibits primarily *s/s'*-symmetry. This phenomenology is expected within Ginzburg-Landau theory[5,13,14] when a charge density wave[5,27] with *d*-symmetry form factor[28-30] and wavevector $Q_C = Q_P$ coexists with a homogeneous *d*-symmetry superconductor; it is also encompassed by several contemporary microscopic theories for the pseudogap phase[18-21].**

*1*     When hole doping suppresses the cuprate antiferromagnetic insulator state the pseudogap regime emerges[4]. Both the superconducting state (SC) and a charge density wave (CDW) state[27] exist therein. However, a "pair density wave" state, in which the Cooper-pair density modulates spatially at wavevector ***Q***, is also widely hypothesized[5-21] to exist in the pseudogap regime. There are compelling theoretical motivations for a cuprate PDW state. First, microscopic theories for local electronic structure of the hole-doped antiferromagnet predict a cuprate PDW state linked to combined modulations of hole-density and antiferromagnetic spin-density[6-10]. Second, *c*-axis superconductivity does not appear in La$_{2-x}$Ba$_x$CuO$_4$ until temperatures far below where the CuO$_2$ planes superconduct, a situation that could be explained by having an orthogonal PDW state in each CuO$_2$ plane[5,11,13]. Third, a plausible explanation exists for the unusual characteristics of single-particle excitations if a cuprate PDW state occurs[15-17]. Finally, theories for the cuprate pseudogap phase hypothesize a composite state in which the CDW is linked by an SU(2) rotation to a PDW of the same wavevector[18-21].

*2*     The definitive signature[1,2,5] of a PDW would be a periodic modulation of the Cooper-pair condensate density with wavevector ***Q***. In principle, this could be visualized using two techniques. First, scanning tunneling microscopy (STM) using conventional single-electron tunneling may detect modulations at ***Q*** in the locally defined energy gap

$$\Delta(\boldsymbol{r}) = \Delta_0 + \Delta_P cos(\boldsymbol{Q} \cdot \boldsymbol{r}) \tag{1a}$$

Here $\Delta_0$ is the homogenous superconductor energy gap and $\Delta_P cos(\boldsymbol{Q} \cdot \boldsymbol{r})$ that of the PDW (Refs. 1,2). A second possibility is scanned (Josephson) tunneling of Cooper-pairs[22-24] to detect the modulations of Cooper-pair density directly. Again, in the presence of the homogenous condensate, this should result in a modulating Josephson critical current $I_J$

$$I_J(\boldsymbol{r}) = I_J^0 + I_J^P cos(\boldsymbol{Q} \cdot \boldsymbol{r}) \qquad (1b)$$

where the first term represents Cooper-pair tunneling to the homogenous condensate and the second to the PDW. Imaging the spatial arrangements of superfluid density $\rho_S(\boldsymbol{r})$ would then be possible, since $I_J^2(\boldsymbol{r}) \propto \rho_S(\boldsymbol{r})$ (Methods I).

**3** The phenomena in Eqn. 1 have not been detected in a cuprate (or any other) superconductor. One challenge is the extreme variation in d$I$/d$V$($E$=e$V$) tunneling spectra (Fig. 1b) typically detected at the $Bi_2Sr_2CaCu_2O_8$ surface $T(\boldsymbol{r})$ (Fig. 1a). Visualizing the maximum energy gap (black arrows Fig. 1b) yields a heterogeneous 'gapmap' $\Delta(\boldsymbol{r})$ (Fig. 1c), whose Fourier transform $\tilde{\Delta}(\boldsymbol{q})$ (Fig. 1d) exhibits only one sharp finite-$\boldsymbol{q}$ peak (blue arrow). This is the $\Delta(\boldsymbol{r})$ modulation at wavevector $\boldsymbol{Q}_{SM}$ caused by a quasi-periodic distortion of the $Bi_2Sr_2CaCu_2O_8$ crystal unit cell along the $CuO_2$ (1,1) direction[26] (Fig. 1e). As no other periodic modulations of the maximum energy gap $\Delta$ (or any other energy gap) have been detected in cuprate gapmap studies, searches for a cuprate PDW via Eqn. 1a yield a null result.

**4** Use of superconducting STM tips to measure the spatial variation of Cooper-pair tunneling (SJTM) then remains the most promising approach to search for a PDW state (Eqn. 1b). For two superconducting electrodes with energy gaps $\Delta(T)$ and phases $\phi_1$ and $\phi_2$, the Josephson current is given by $I(\phi) = I_J sin(\phi)$ where $\phi = \phi_2 - \phi_1$. The Josephson critical current $I_J$ is given by $I_J R_N = \frac{\pi \Delta(T)}{2e} tanh(\Delta(T)/2kT)$ where $R_N$ is the normal state junction resistance, $k$ the Boltzmann constant, and 2e the Cooper pair charge. Time independence of $\phi$ would require the Josephson energy $E_J = {\phi_0 I_J}/{2\pi}$, where $\phi_0$ is the magnetic flux quantum, to exceed the thermal fluctuation energy $kT$. For a nanometer dimension Josephson junction with sub-nanoamp $I_J$, this regime remains out of reach at sub-millikelvin temperatures. However, imaging of phase-diffusion dominated Josephson tunneling is possible by using superconducting-tip STM to measure an $I(V)$ whose maximum current $I_c \propto I_J^2$ (Ref. 22-24 and Methods I).

**5**   To search for a cuprate PDW near $\boldsymbol{Q} \approx (0.25,0)2\pi/a_0 ; (0,0.25)2\pi/a_0$ where $a_0 \approx$ 3.8 Å also requires visualizing $I_c(\boldsymbol{r}) \propto I_J^2(\boldsymbol{r})$ with nanometer resolution. Moreover, high $\Delta$, low $R_N$ and millikelvin operating temperatures would all be highly advantageous. Motivated thus, we use a dilution refrigerator based STM operated below 50 mK, and achieve a high tip-energy-gap $\Delta_T$ by picking up a nanometer sized flake of Bi$_2$Sr$_2$CaCu$_2$O$_8$ on the end of our tungsten tip (Fig. 2a and Methods II). This immediately converts our measured NIS tunneling spectrum (dashed line Fig. 2b) to a SIS spectrum (solid line Fig. 2b) exhibiting energy separation $\approx 2(\Delta_S + \Delta_T)$ between conductance peaks (Methods II). Figure 2b demonstrates that our $d$-wave HTS tip has $\Delta_T \approx 25$ meV (double headed arrows), and its spatial resolution is evaluated as ~1nm using a topographic image of the BiO surface of Bi$_2$Sr$_2$CaCu$_2$O$_8$ (Fig. 2c) measured using this tip in the single-particle tunneling regime (Methods II). Next, we measure the $I(V)$ characteristic of this tip as a function of decreasing tip distance from the bulk crystal surface, and thus decreasing $R_N$. At $T$=45 mK and fixed $\boldsymbol{r}$, as the SC tip is moved forward in ~10 pm steps the evolution of $I(V)$ characteristics is as shown in Fig. 2d. As the Josephson $I_J$ increases with diminishing distance, the maximum observable current $I_c \propto I_J^2$ increases as expected (Methods I and II).

**6**   To directly test the Cooper-pair condensate visualization capabilities of this $I_c(\boldsymbol{r})$ imaging technique, we use two approaches. First, when Zn impurity atoms are substituted at the Cu sites, muon spin rotation studies show that the Cooper-pair condensate is completely suppressed in a few nm-radius area around each Zn, in a "swiss cheese" configuration[25]. Figure 3a shows an 76 nm X 76 nm image of SIS single particle tunneling conductance $g(\mathbf{r},-20\text{ meV})$ using the same tip. One can locate each Zn impurity atom by its scattering resonance peak that is shifted by convolution with the HTS tip spectrum from its resonance energy $\Omega$=-1.5 meV to $|E|\sim|\Omega|+|\Delta_T|\sim 25$ meV (vertical red line Fig. 3a). Thus, the Zn impurity atoms are at the dark spots in Fig. 3a. In Fig. 3b we show $I_c(\mathbf{r})$ measured at ~50 mK in the identical FOV as Fig. 3a. This reveals how $I_c(\mathbf{r})$, and thus the Cooper-pair condensate, is suppressed within a radius $R\sim 3$ nm of each Zn atom site consistent with muon spin rotation studies[25]. A second test is possible because of the crystal "supermodulation" and associated modulating superconductivity at wavevector $\mathbf{Q}_{SM}$ [26] (Fig. 1f). In Fig. 3c we plot the magnitude of the Fourier transform $\tilde{T}(\mathbf{q})$ in grey and $\tilde{I}_c(\mathbf{q})$ in orange, both measured simultaneously along the (1,1) direction through $\mathbf{Q}_{SM}$. This demonstrates directly the capability to visualize the Cooper-pair density modulations induced by the Bi$_2$Sr$_2$CaCu$_2$O$_8$ crystal-supermodulation. Overall, these tests show that nm-resolution Cooper-pair condensate visualization is achievable using this *d*-wave high-gap tip, plus millikelvin operation SJTM approach.

**7** Next we apply this technique to search for a cuprate PDW in the same Bi$_2$Sr$_2$CaCu$_2$O$_8$ samples with $T_c$= 88K and $p$=17 % and at $T$<50 mK (Methods IV). The two states already reported to coexist at this *p* are the high-temperature superconductor and a charge density wave with $\mathbf{Q} = (0.22 \pm 0.02,0)2\pi/a_0; (0,0.22 \pm 0.02)2\pi/a_0$ (Ref. 27). Figure 4a shows our measured $I_c(\mathbf{r})$ in the 35nmX35nm FOV (dashed box Fig. 3b). Clear $I_c(\mathbf{r})$ modulations are immediately observable and, as $I_c(\mathbf{r}) \propto \left(I_J^0\right)^2 + 2I_J^0 I_J^P \cos(\mathbf{Q}.\mathbf{r}) + \left(I_J^P\right)^2 \cos^2(\mathbf{Q}.\mathbf{r})$ from Eqn. 1b, they provide the magnitude and wavevector of component $I_j^P(r)$ modulations. The magnitude of the Fourier transform of $I_c(r)$, $|\widetilde{I}_c(q)|$, is shown in Fig. 4b revealing that modulation in $I_c(r)$ occurs at the wavevectors $\boldsymbol{Q}_P = (0.25 \pm 0.02, 0)2\pi/a_0; (0,0.25 \pm 0.02)2\pi/a_0$ (dashed red circles Fig. 4b), while nothing is observable in the second harmonic $cos^2(\boldsymbol{Q}.\boldsymbol{r})$. This situation occurs because the $I_j^P(r)$ modulations are superposed on a much stronger spatially constant critical current $I_j^0$ (Eqn. 1b). These data provide strong *a priori* evidence for the existence of a pair density wave coexisting with a robust homogenous Cooper-pair condensate in Bi$_2$Sr$_2$CaCu$_2$O$_8$. In Fig. 4c we show as blue dots the measured value of $I_c(r)$ along the fine blue line in Fig. 4a, while as a fine red line we show the amplitude and wavelength of the global $I_c(r)$ modulations (determined from the magnitude and central $\boldsymbol{Q}$ value of the peaks in Fig. 4b). From these data, and in general from the magnitude of the peaks at $\boldsymbol{Q}_x$ and $\boldsymbol{Q}_y$ in Figs. 4b, we conclude that the modulation amplitude in Cooper-pair density is ~5% of its homogeneous value. Therefore the cuprate PDW at *p*=17% is quite subdominant to the homogeneous *d*-wave superconducting state.

**8** Finally, in Fig. 4d we study $\widetilde{D}(\boldsymbol{q}) = \widetilde{O}_x(\boldsymbol{q}) - \widetilde{O}_y(\boldsymbol{q})$ the oxygen-sublattice-phase-resolved image of *d*-symmetry form factor density modulations[28-30], from a Bi$_2$Sr$_2$CaCu$_2$O$_8$ sample with the same hole-density (Methods VI). The conventional locations of *d*-symmetry form factor CDW peaks in $\widetilde{D}(\boldsymbol{q})$ (dashed red circles) occur at $\boldsymbol{Q}_C = (0.22 \pm 0.04, 0)2\pi/a_0; (0,0.22 \pm 0.04)2\pi/a_0$. Note that in the unprocessed data the modulations in electronic structure and topography occur based upon the sum of both oxygen sublattices ($\widetilde{O}_x(\boldsymbol{q}) + \widetilde{O}_y(\boldsymbol{q})$) and exhibit $\boldsymbol{Q} = \boldsymbol{Q}_{BRAGG} \pm \boldsymbol{Q}_C$ (Ref. 28). No response of the tip to these modulations could produce a spurious $I_c(r)$ modulation at the PDW wavevector $\boldsymbol{Q}_P = (0.25 \pm 0.02, 0)2\pi/a_0; (0,0.25 \pm 0.02)2\pi/a_0$ (Methods VI). Obviously, Fig. 4b and Fig 4d are not identical, with the PDW exhibiting significantly narrower peaks and thus more spatial coherence. Nevertheless, this PDW wavevector (Fig. 4b, c) is not inconsistent (within joint error bars) with a conventionally-defined wavevector of the cuprate CDW state at the same hole density[27]. However, this PDW also exhibits a primarily *s/s'*-symmetry form factor because Fig. 4b is based on the conventional sum over sublattices $\tilde{O}_x(\boldsymbol{q}) + \tilde{O}_y(\boldsymbol{q}) + \widetilde{Cu}(\boldsymbol{q})$, while the CDW state exhibits a primarily *d*-symmetry form factor as detected in the difference of sublattice phase resolved images[28] $\tilde{O}_x(\boldsymbol{q}) - \tilde{O}_y(\boldsymbol{q})$ (Fig. 4d).

**9**     Visualization of a Cooper-pair density wave, a long-term challenge for physics[1,2], has now been achieved using scanned Josephson tunneling microscopy[22-24]. The subdominant PDW observed in $Bi_2Sr_2CaCu_2O_{8+x}$ at approximately the same wavevector as a CDW ($\boldsymbol{Q_P} \approx \boldsymbol{Q_C}$) and with *s/s'* form factor (Fig. 4), is as expected within Ginzburg-Landau theory[5,11,13,14] in the case (among others) where a *d*-symmetry superconductor coexists with a *d*-symmetry form factor CDW (Methods VII). Microscopic theories for a cuprate PDW[5-9,13,16] requiring $\boldsymbol{Q_P} = \boldsymbol{Q_C}/2$ cannot be fully tested here because of disorder in $\tilde{I}_c(\boldsymbol{q})$ at low $\boldsymbol{q}$ (Fig. 4b). However, microscopic models for the pseudogap phase involving interplay of *d*-symmetry Cooper pairing and a *d*-symmetry form factor CDW do yield a PDW with $\boldsymbol{Q_P} = \boldsymbol{Q_C}$ [18-21] as in Fig. 4, although typically with a *d*-symmetry form factor. Finally, the HTS-tip millikelvin-SJTM approach to Cooper-pair condensate imaging introduced here provides a new and direct route to visualization of FFLO[1,2] or PDW[6-21] states in other cuprates, iron-pnictides and unconventional superconductors.

# Figure Captions

**Figure 1 Spatial Variations and Modulations in Cuprate Energy Gaps**

a. Typical 35 nm X 35 nm topographic image $T(r)$ at BiO termination layer of $Bi_2Sr_2CaCu_2O_8$ (crystal "supermodulation" runs vertically).
b. Typical $g(E)=dI/dV(E=eV)$ differential tunnel conductance spectra of superconducting $Bi_2Sr_2CaCu_2O_8$. The maximum energy gap $\Delta$ is determined from half the distance between peaks in each spectrum.
c. Spatial arrangement of $\Delta(r)$ (gapmap) for $p\sim17$ % $Bi_2Sr_2CaCu_2O_8$ samples studied here in same 35 nm X 35 nm FOV as a.
d. Magnitude of Fourier transform of c, $|\tilde{\Delta}(q)|$ (crosses are at $q=(\pi/a_0,0);(0, \pi/a_0)$) As typical[26], a single inequivalent peak due to the crystal "supermodulation" is observed (blue arrow).
e. $|\tilde{T}(q)|$, magnitude of the Fourier transform of a (crosses are at $q=(\pi/a_0,0);(0, \pi/a_0)$). The "supermodulation" is a quasi-periodic modulation in unit cell dimensions along the (1,1) direction of the $CuO_2$ plane. Its wavevector $Q_{SM}$ is indicate by blue arrow.
f. Simultaneously measured magnitude of $\tilde{\Delta}(q)$ and $\tilde{T}(q)$ from d,e along the (1,1) direction. Their primary peaks coincide exactly.

**Figure 2 d-wave HTS Tip Fabrication for SJTM**

a. Schematic of tungsten STM tip with nanometer $Bi_2Sr_2CaCu_2O_8$ flake adhering. Inset: SI-STM and Josephson circuitry used where $R_B$ =10 M$\Omega$.
b. Conversion of single-particle NIS (dashed) to single-particle SIS (solid) spectra when $Bi_2Sr_2CaCu_2O_8$ nano-flake adheres to W tip. These data are from same experiment described throughout this manuscript but multiple such d-wave HTS tips have been created (Methods II). Tip gap $\Delta_T \approx 25$ meV is the difference between NIS and SIS peaks (red arrows).

c. Topographic image (76 nm X 76 nm) of BiO termination surface using single-particle SIS tunneling with the same tip as in a,b. Spatial resolution is ~1 nm. Absence of Moiré patterns in its Fourier transform (Methods II) reveals that nano-flake crystal axes (order parameter) are aligned with the axes (order parameter) of the bulk crystal.

d. Measured evolution of the $I(V)$ characteristic with diminishing tip/sample distance and thus diminishing $R_N$, at $T=45$ mK (junction formation conditions in blue). The maximum current $I_c$ for a typical $I(V)$ curve is indicated by a green arrow;

**Figure 3 Cooper Pair Condensate Visualization using SJTM**

a. Convolution of $\Omega=-1.5$ meV resonance at each Zn impurity atom with the d-wave spectrum of the tip (dashed curve Fig. 2a) shifts resonance signature to $|E|\sim|\Omega|+|\Delta_T|$. Measured SIS $g(r,E=-20$ meV) 76 nm X 76 nm image of $Bi_2Sr_2CaCu_2O_8$ near this energy (vertical red line) identifies site of each Zn.

b. A 76 nm X 76 nm $I_c(r)$ map measured at 45mK in same FOV as a. A deep minimum (~95% suppression) occurs in $I_c(r)$ surrounding each Zn site where the Cooper-pair condensate is suppressed[25].

c. Simultaneously measured magnitude of $\tilde{I}_c(q)$ and $\tilde{T}(q)$ along the $CuO_2$ (1,1) direction. The primary peak in $\tilde{I}_c(q)$ coincides exactly with that of the crystal "supermodulation" demonstrating visualization of Cooper-pair density modulations.

**Figure 4 Visualizing the Cooper-Pair Density Wave in $Bi_2Sr_2CaCu_2O_8$**

a. Scanned $I_c(r)$ image in 35 nm X 35 nm FOV (dashed box Fig. 3b). Periodic modulations in $I_c(r)$ along (1,1) due merely to crystal supermodulation have been removed. Modulations in $I_c(r)$ along the $x,y$ axes of the in-plane CuO bonds, are seen clearly and directly.

b. Magnitude of Fourier transform of $I_c(r)$ in a, $|\tilde{I}_c(q)|$ (crosses at $q=(\pi/a_0,0);(0, \pi/a_0)$). Maxima due to modulations in $I_c(r)$ (dashed red circles) occur at $Q_P=(0.25,0)2\pi/a_0;(0,0.25)2\pi/a_0$. No significant modulations occur in $\tilde{R}_N(Q_P)$ ( Methods V).

c. Measured values of $I_c(r)$ along the dashed blue line in a (blue dots); statistical error bars are defined as variance of $I_c$ transverse to dashed line. Fine red line shows the global amplitude and $Q_P$ of modulations in $I_c(r)$, as determined from the magnitude and central $Q_P$ value of the maxima in b. Amplitude of the $I_c(r)$ modulations is about 5% of the mean value of $I_c(r)$.

d. Magnitude of Fourier transform of $\tilde{D}(q) = \tilde{O}_x(q) - \tilde{O}_y(q)$ revealing the d-symmetry form factor density wave measure at the same hole-density as b (crosses at $q=(\pi/a_0,0);(0, \pi/a_0)$). Density wave modulations in $|\tilde{D}(q)|$ (dashed red circles) occur at $Q_C=(0.22,0)\ 2\pi/a_0;(0,0.22)2\pi/a_0$ consistent with Ref. 27.

## Acknowledgements

We acknowledge and thank D. Agterberg, A.V. Balatsky, D. Chowdhury, A. Chubukov, E. Fradkin, R. Hulet, S.A. Kivelson, P.A. Lee, M. Norman, J.W. Orenstein, C. Pepin, S. Sachdev, J. Tranquada, and Y. Wang for very helpful discussions, advice and communications. Development and operation of HTS SJTM technology and M.H.H and A.K., were funded by the Moore Foundation's EPiQS Initiative through Grant GBMF4544; SDE acknowledges studentship funding from the EPSRC under Grant EP/G03673X/1; J.C.D & A.P.M. acknowledge research support from EPSRC through the Programme Grant '*Topological Protection and Non-Equilibrium States in Correlated Electron Systems*'. S.U. and H.E. acknowledge support from a Grant-in-Aid for Scientific Research from the Ministry of Science and Education (Japan); S.H.J. and J.L. acknowledge support from the Institute for Basic Science, Korea under IBS-R009-D1; J.C.D and K.F. acknowledge salary support from the U.S. Department of Energy, Office of Basic Energy Sciences, under contract number DEAC02-98CH10886; E.A.K. acknowledges support by the U.S. Department of Energy, Office of Basic Energy Sciences, Division of Materials Science and Engineering under Award DE-SC0010313.

**Author Contributions:** M.H.H., S.D.E., A.K., and J.L. developed the SJTM techniques and carried out the experiments; K.F., H.E. and S.U. synthesized and characterized the samples; M.H.H., S.D.E., A.K., S.H.J and K.F. developed and carried out analysis; E.-A.K. and M.L. provided theoretical guidance; A.P.M., J.L. and J.C.D. supervised the project and wrote the paper with key contributions from M.H.H., S.D.E. and K.F. The manuscript reflects the contributions and ideas of all authors.

\* To whom correspondence should be addressed: jinholee@snu.ac.kr and jcseamusdavis@gmail.com

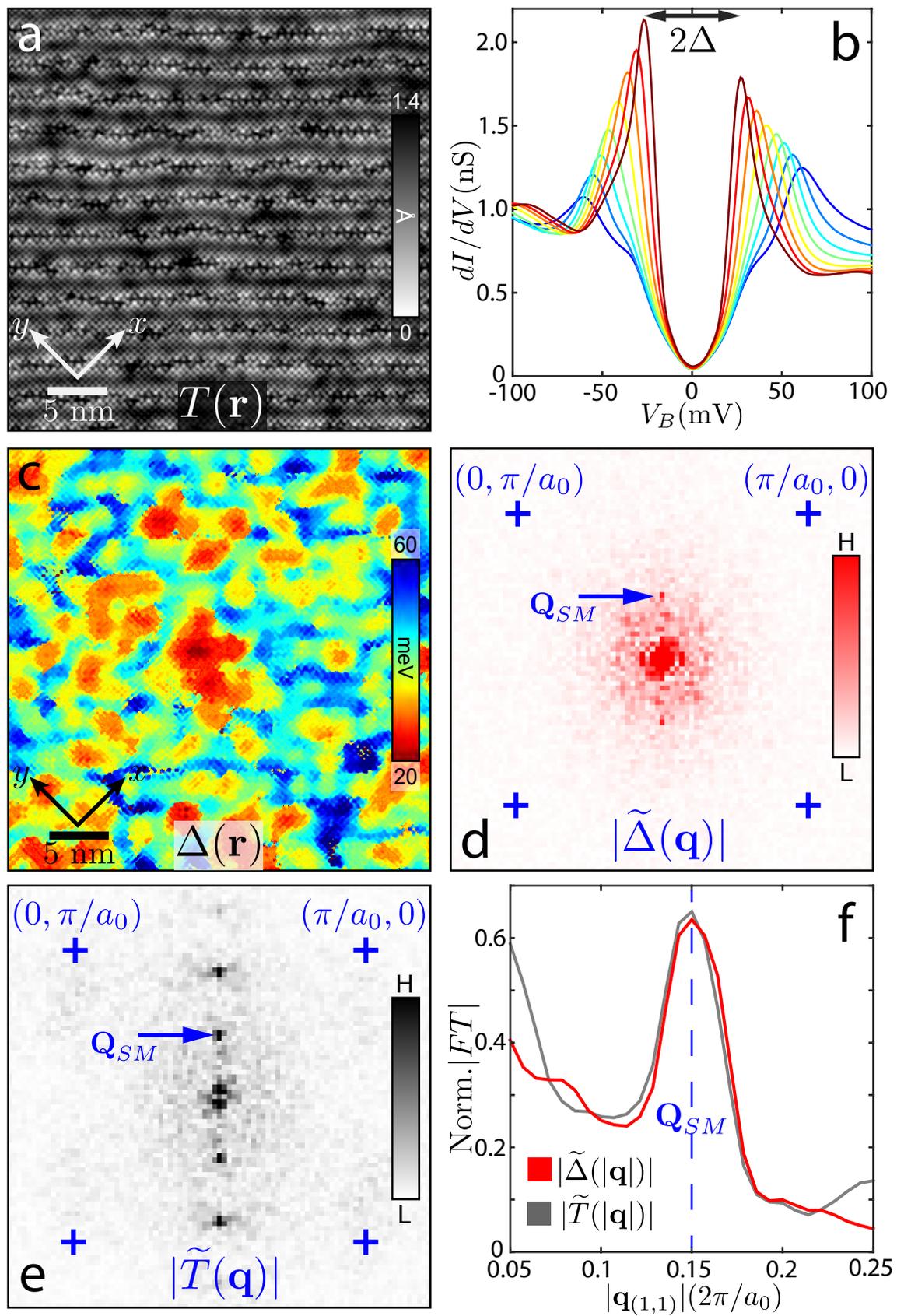

FIG2

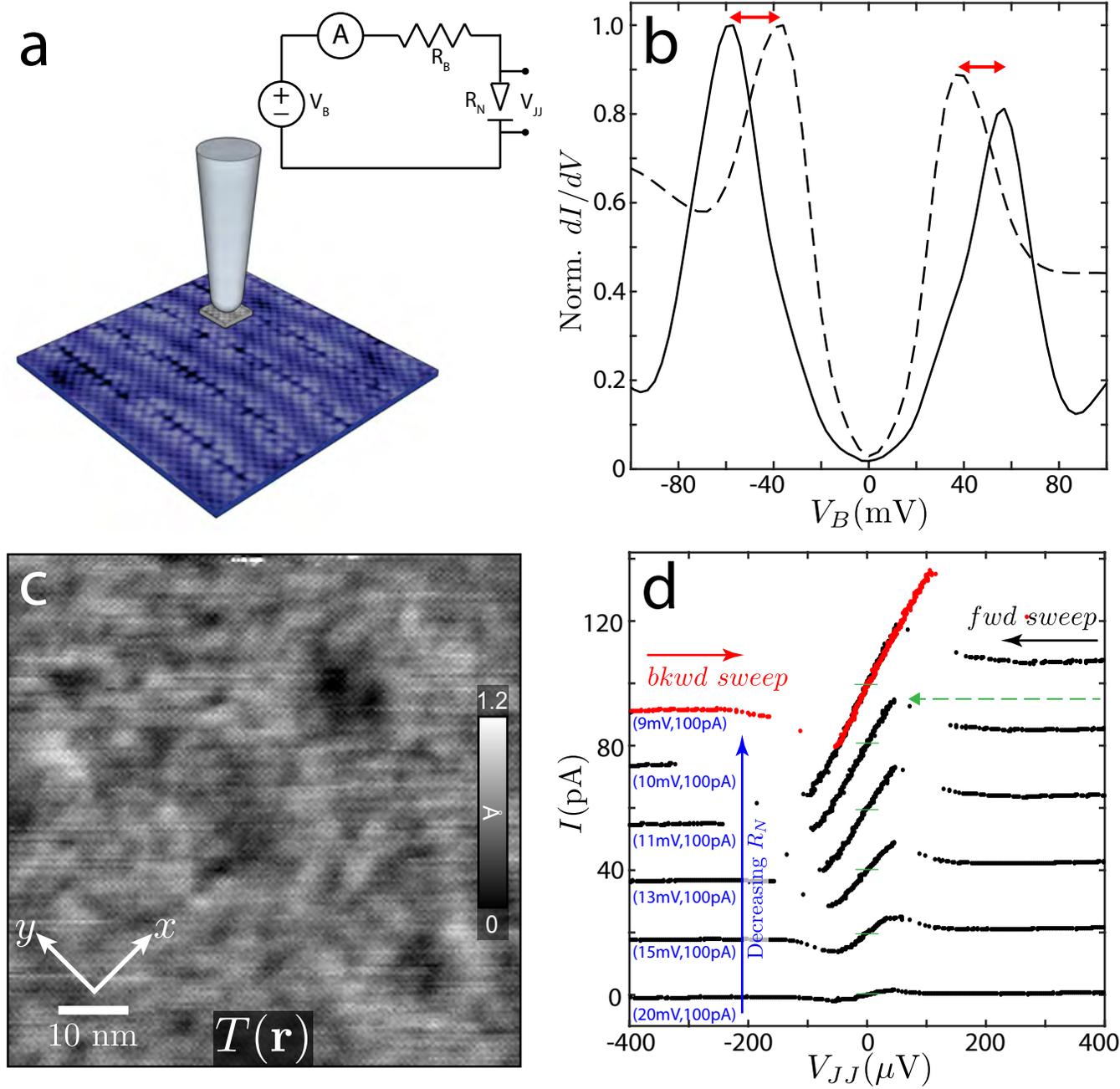

FIG3

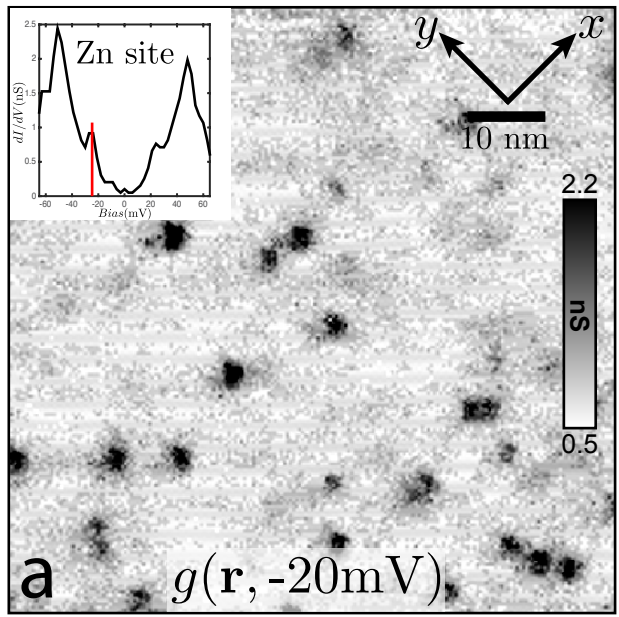

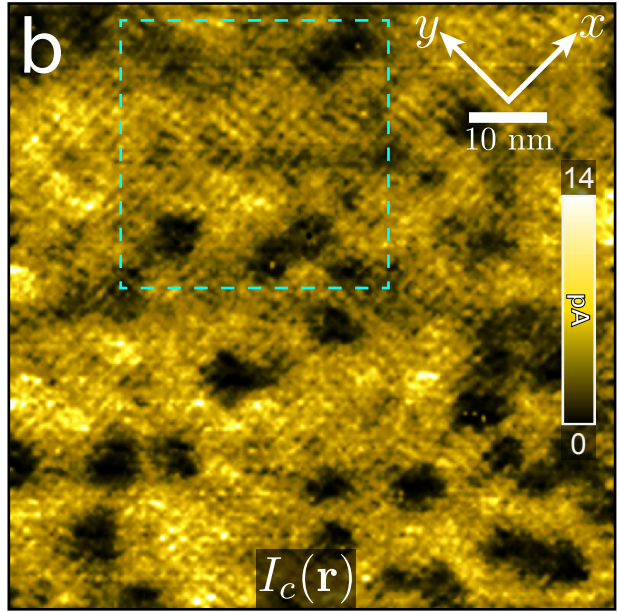

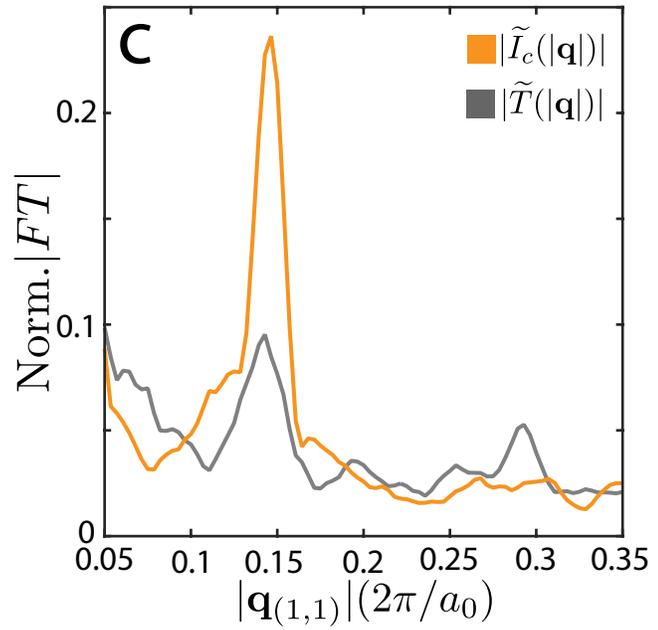

FIG4

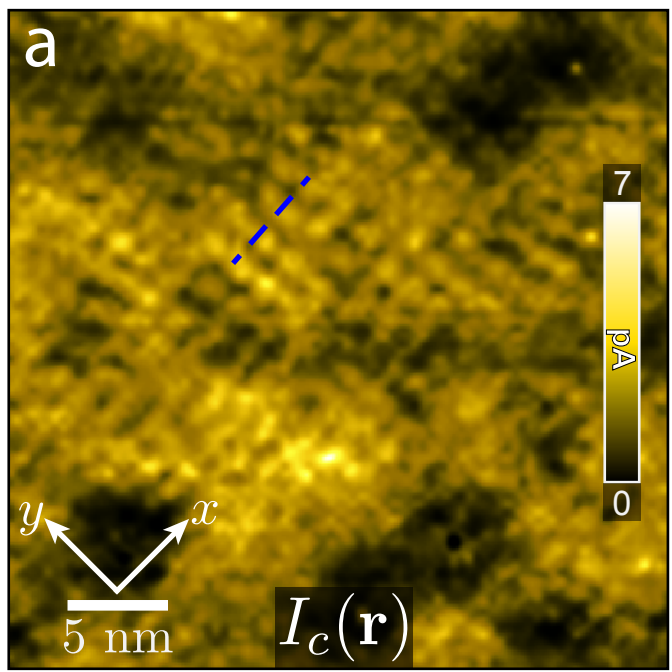
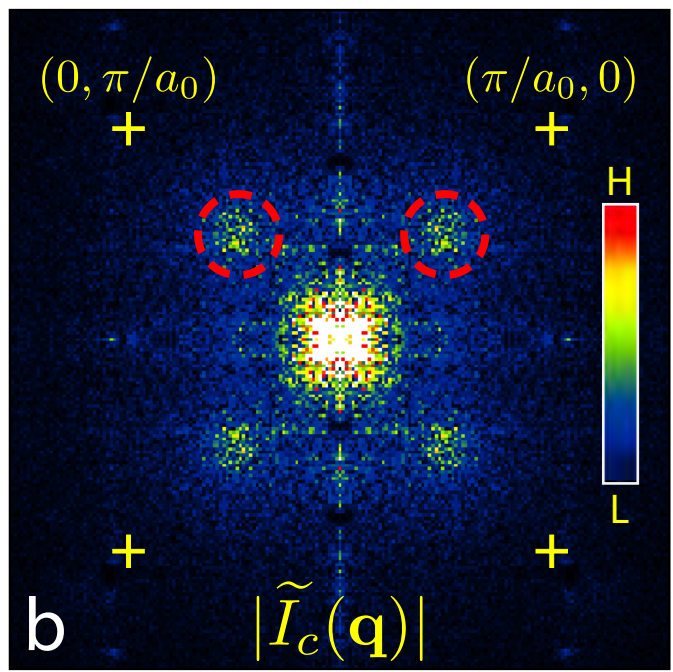
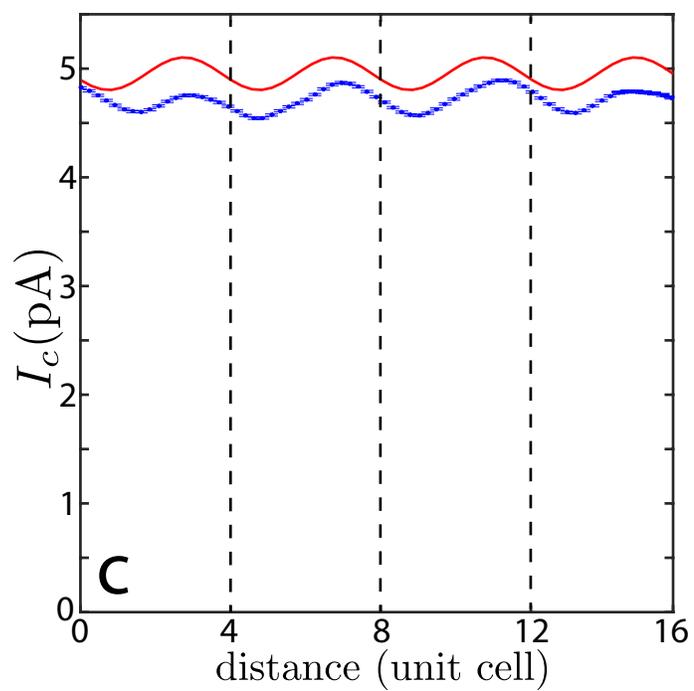
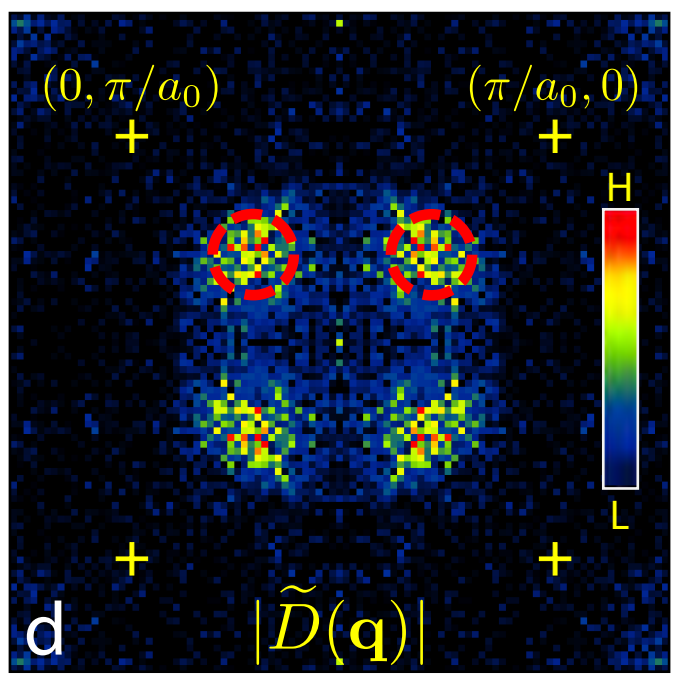

## Methods

### (I) Modeling Phase Diffusion Josephson Dynamics for SJTM

The Josephson effect[31] can be described by two fundamental equations. The first
$$I = I_J sin(\phi) \qquad \text{M1.1}$$
describes the relationship of the Cooper-pair current $I$ flowing through the junction to the phase difference $\phi$ between superconducting order parameters in the two electrodes, with $I_J$ the Josephson critical current. The second
$$\frac{d\phi}{dt} = \frac{2e}{\hbar}V \qquad \text{M1.2}$$
relates the time evolution of that phase difference to the voltage dropped across the junction. In cases where the Josephson coupling energy $E_J = \frac{\hbar}{2e}I_J$ is much larger than $kT$, where $k$ is the Boltzmann constant, and the effect of thermal fluctuations in $\phi$ can be ignored, there exists a steady state solution to equations M1.1 and M1.2 where $V=0$ for $I < I_J$. At 50 mK, the condition $E_J = kT$ corresponds to a Josephson critical current $I_J \sim 2$ nA.

Because of our operating junction resistance $R_N \sim 10$ MΩ, the maximum Josephson current used in the SJTM experiments described here was approximately $\sim 20$ pA, with no zero resistance state being detected. Thus, our experimental junction lies in a regime where $\phi$ exhibits steady-state diffusion through the Josephson potential due to thermal fluctuations[32-37]. This results in a voltage dropped across the junction while the Cooper-pair current is flowing. We analyze the Josephson $I(V)$ characteristics between our $d$-wave HTS tip and the Bi$_2$Sr$_2$CaCu$_2$O$_8$ surface using this classical phase-diffusion model for small junctions[32,33] within which the $I(V)$ characteristic of the junction is given by

$$I(V_{JJ}) = \frac{1}{2}I_J^2 Z \frac{V_{JJ}}{V_{JJ}^2 + V_c^2} \qquad \text{M1.3}$$

$V_{JJ}$ is the voltage across the junction and $V_c = \frac{2eZkT^*}{\hbar}$ with $Z$ the impedance in series with the voltage source at the high frequencies relevant to repeated re-trapping of the phase involved in diffusion. Importantly, $I(V_{JJ})$ is maximal at

$$I_c = \frac{\hbar}{8ekT^*} I_J^2 \qquad \text{M1.4}$$

where $T^*$ is an effective temperature parameterizing how dissipative the phase evolution is. Spatially resolved measurements of $I_c(\boldsymbol{r})$ thus allow one to measure spatial modulations in $I_J^2(\boldsymbol{r})$ (Refs 22-24).

Extended Data Fig. 1a shows a diagram of the hybrid SISTM/SJTM measurement circuit used in the experiments presented here. A load resistor $R_B$ = 10 MΩ is voltage biased in series with the Josephson junction formed between STM-tip and sample. Extended Data Fig. 1b shows the DC solutions to an electronic circuit model for this setup as shown in Extended Data Fig. 1a. The red and blue lines show the trajectories that the forwards and backwards sweeps take through the $I(V)$ plane. The non-linearity of the circuit introduces discontinuities in the measured $I(V)$; a discontinuity occurs at the point where $I=I_c$ and thus makes this current experimentally identifiable due to the sharp $I(V)$ feature associated with it. Extended Data Figure 1c shows these same trajectories but as a function of the voltage across the junction $V_{JJ}$. Another result of this circuit's non-linearity is that hysteresis should occur in the $I(V)$ characteristic even if the dynamics of the Josephson junction itself are over-damped at all frequencies. This hysteresis is indeed observed throughout our SJTM studies of Bi$_2$Sr$_2$CaCu$_2$O$_8$ (main text Fig. 2d), but systematic errors due to it are avoided during our $I_c(\boldsymbol{r})$ imaging experiments by sweeping the applied voltage always in the same direction once the Josephson junction has been formed at each location $\boldsymbol{r}$.

**(II) Characterizing Bi$_2$Sr$_2$CaCuO$_8$ Nano-flake STM Tips**

To detect the gap modulations described by Eqn. 1a requires single electron tunneling from a non-superconducting electrode (NIS), while to observe the Josephson current modulations in Eqn. 1b requires Cooper-pair tunneling from a superconducting electrode (SIS). For our SJTM studies, the high energy-gap at the STM tip $\Delta_T$ is achieved by picking up a nanometer sized flake of Bi$_2$Sr$_2$CaCu$_2$O$_8$ (BSCCO) onto the conventional W microscope tip. Extended Data Figure 2 shows the measured d$I$/d$V$ spectrum of the W tip before that process as a dashed line; it is precisely as expected for hole-density $p\sim0.17$ Bi$_2$Sr$_2$CaCu$_2$O$_8$ samples. By using the standard tunneling equation for two electrodes with density of states $DOS(E)$, each identical to the dashed line, one can simulate the expected d$I$/d$V$ from a BSCCO nano-flake tip. The solid line is such a simulation and is in very good agreement with the typical SIS spectrum shown in Fig. 2b and observed throughout these studies.

In Extended Data Fig. 3 we show a comparison of performance of two completely different BSCCO nano-flake tips, assembled and used during two completely different experimental SJTM studies. Extended Data Figures 3a,b show the measured d$I$/d$V$ spectra of the two distinct BSCCO nano-flake tips; they are obviously highly similar. Extended Data Figure 3c,d demonstrate the very similar $\sim 1$ nm spatial resolution in topographic images made by using these two tips operating with the same circuitry in Extended Data Fig. 1. Extended Data Figure 3e,f show the Fourier transform of these two topographic images, demonstrating that the supermodulation is easily detected in both cases and that no other unusual characteristics, such as additional $\boldsymbol{q}$-space peaks or Moiré patterns, are observed.

To evaluate relative orientation of the BSCCO-nano flake crystal axes to the axes of the sample studied, we evaluate simulations for what is expected of such a tip. Extended Data Figures 4a,b show a measured SIN topograph $T(\boldsymbol{r})$ of BSCCO taken with a conventional W tip, and its Fourier transform. To simulate what would be observed if the crystal axes of a BSCCO nano-flake are (not) oriented to those of the sample crystal axes we extract a typical $\sim 2$ nm diameter patch from this image and then convolute the patch image with $T(\boldsymbol{r})$ at each location $\boldsymbol{r}$. This results in the simulated topograph expected when using a BSCCO nano-flake tip with (without) axes aligned to the crystal axes. If the axes of the patch are aligned (Extended Data Fig. 4c) with those of the crystal the results is shown in Extended Data Fig. 4e (Fourier transform inset). If the axes of the patch are misaligned (Extended Data Fig. 4d) with those of the crystal the results is shown in Extended Data Fig. 4f (Fourier transform inset). As the vivid Moiré effects seen in the Fourier transform of the latter simulation are not observed in the actual experiments (Extended Data Fig. 3e,f), we conclude that the axes of the nano-flake are aligned with those of the BSCCO bulk crystal that is being studied.

**(III) Specifications for HTS-Tip, mK-SJTM for Condensate Visualization**

The technical challenge of large field-of-view nanometer resolution SJTM imaging is to measure, in a reasonable time (days), an array of $I_c(\boldsymbol{r})$ values with $\sim 1$ % precision using typically 256 X 256 pixels. Four basic elements are required: (1) an ultra-low-vibration dilution-refrigerator-based STM that is engineered for sub-picometer stability in taking spectroscopic data while out of feedback, and typically at $T$=50 mK; (2) a high temperature superconducting tip creation scheme which involves removing a nanometer sized flake from the BSCCO surface and adhering it to the tip; (3) modification to the electronic circuitry to form a hybrid SISTM/SJTM operating condition (Fig. 2a) which allows conventional operation at the sub-volt bias range in the SIS configuration for topographic imaging and motion control, and bias voltage sweeps in the micro-volt range when in Josephson mode; and (4) control capability to reliably and repeatedly carry out spectroscopic maps $I(V, \boldsymbol{r})$ using a junction resistance $\sim 10$ MΩ at every $\boldsymbol{r}$ and with $V$ spanning tens of μV.

**(IV) Defining, Measuring and Imaging $I_c(\boldsymbol{r})$**

Extended Data Figure 5a shows the histogram of $I_c$ values measured in a typical field of view during these studies. By averaging over the individual measured $I(V)$ from all the locations $\boldsymbol{r}$ contained within the bins indicated by the colored arrows, we obtain the three representative Josephson $I(V)$ measurements shown in Extended Data Figure 5b. They span the range of $I(V)$ characteristics measured at different locations. As an example of the procedure to establish $I_c$, the colored arrows in Extended Data Figure 5b show where $I_c$ is identified in a particular curve. In general we functionally define the quantity $I_c(\boldsymbol{r})$ reported in the main text to be the maximum magnitude of current achieved on sweeping away from $V$=0. By comparison with Extended Data Fig. 1b one can see that this current is the maximum current, $I_c$, described by the phase diffusion model (Methods I) in which, at constant temperature, $I_c(\boldsymbol{r})$ is directly proportional to $I_J^2(\boldsymbol{r})$.

Extended Data Fig. 6a,b,c show typical examples of repeated $I_c(\boldsymbol{r})$ measurement experiments (with slightly different operating parameters but same BSCCO nano-flake tip). The functionality and repeatability of this technique to achieve $I_c(\boldsymbol{r})$ imaging with nanometer resolution is clear.

The typical evolution of measured $I(V)$ during SJTM imaging of $I_c(\boldsymbol{r})$ at 45mK at a specific sequence of locations (numbered 1-9 in Extended Data Fig. 6b) is shown in Extended Data Fig. 7. The almost complete (~95 %) suppression of $I_c$ as this sequence passes through the site of a Zn atom is seen in panel 5. These typical data demonstrate directly the unprocessed signal-to noise for measurement of $I_c$ and how spatial evolution of unprocessed $I(V)$ spectra yields the high fidelity in $I_c(\boldsymbol{r})$ images.

In Extended Data Fig. 8a,b we show the topographic images taken with the same BSCCO nano-flake tip before/after a typical $I_c(\boldsymbol{r})$ mapping experiment. Comparison of the topographs shows them to be virtually identical and demonstrates that no significant changes occur in the geometry of the tip during the map. The implication is that the nano-flake at the end of the tip never once touches the surface, and that all such $I_c(\mathbf{r})$ studies reported are in the true SIS tunneling limit and very far from making so-called point contact. This comports well with the fact that we establish all the Josephson junctions in $I_c(\mathbf{r})$ maps reported using an effective junction resistance of ~10MΩ for SIS tunneling at approximately $V$=10meV.

Finally, our model is that the BSCCO tip has maximum gap $\Delta_T$ and negligible modulating component, while that of the sample $\Delta_S$ has two components $\Delta_S^0$ (uniform component) and $\Delta_S^P$ (PDW component) and $\Delta_S = \Delta_S^0 + \Delta_S^P$ with $\Delta_S^0 \gg \Delta_S^P$. The Josephson coupling then consists of a primary coupling that occurs between the uniform components, and a secondary coupling that is between the uniform component of the tip and the modulated component of the sample (as Eqn. 1b) with the empirical result of convolution presented in Fig. 4. To leading order $I_c(\mathbf{r})$ is then linear in the PDW component of both the sample gap and superfluid density.

**(V) Visualizing Supermodulation-Induced Pair Density Wave and $R_N(\mathbf{r})$**

In principle SJTM allows direct measurement of modulations in the superconducting order parameter amplitude through the $I_J R_N$ product. Here, $I_J$ is the intrinsic, $T$=0, Josephson critical current of the junction and $R_N$ is a characteristic resistance of the single particle tunneling channel. In taking this product, variations that affect the matrix elements for tunneling of Cooper-pairs and the single-particle tunneling conductance ($1/R_N$) in the same way cancel.

We use the differential conductance of the tunnel junction, at a bias voltage far higher than the SIS gap edge as an estimate of the ungapped, "normal state" junction conductance $G_N$. Simultaneous measurement of this value and $I_c(\mathbf{r})$ were not possible because the former requires application of fractions of a volt while the latter requires bias of a few tens of μV. Moreover, at the tip-sample separations used in measuring $I_c(\mathbf{r})$, the single particle tunneling current can become destructively large to our BSCCO nano-flake tip. Thus the following procedure was adopted:

1) Measure $I_c(\mathbf{r})$ with setpoint current $I_s$=$I_{s1}$ and bias voltage $V_s$=$V_{s1}$ as shown in Extended Data Figure 9a.
2) In the same field of view measure $g_1(\mathbf{r},E)$ defined as $dI/dV$ for bias voltages $V_{s1}$ to $-V_{s1}$ with setpoint condition $I_s$=$I_{s1}$, $V_s$=$V_{s1}$.
3) In the same field of view measure $g_2(\mathbf{r},E)$ defined as $dI/dV$ with setpoint condition

$I_s=I_{s2}$, $V_s=V_{s2}$ for bias voltages $V_{s2}$ to $-V_{s2}$, where $V_{s2} \gg V_{s1}$ and $V_{s2} > \Delta_{SIS}$.

4) Over the range $V = [-V_{s1}, V_{s1}]$ determine the coefficient $\alpha(\mathbf{r})$ which scales $g_2(\mathbf{r})$ onto $g_1(\mathbf{r})$ via $g_1(\mathbf{r},E) = \alpha(\mathbf{r})g_2(\mathbf{r},E) + \beta(\mathbf{r})$. In doing the scaling a small region of voltages near $V=0$ where the Cooper-pair tunneling (Josephson) signal exhibits is excluded.

5) $R_N(\mathbf{r})$ for junctions used to measure $I_c(\mathbf{r})$ is then determined by $1/R_N(\mathbf{r}) = G_N(\mathbf{r}) = \alpha(\mathbf{r})/2\ (g_2(\mathbf{r},+V_{s2}) + g_2(\mathbf{r},-V_{s2})) + \beta(\mathbf{r})$, as shown in Extended Data Fig. 9b.

Now we consider the case of the cuprate supermodulation PDW where the bulk quasi-periodic distortion of the $Bi_2Sr_2CaCu_2O_8$ crystal unit cell at wave-vector $\mathbf{Q}_{SM}$ leads to modulations of the measured $I_c(\mathbf{r})$ at the same wave-vector superposed on the larger translationally invariant component.

$$\Delta(\mathbf{r}) \propto I_J(\mathbf{r})R_N(\mathbf{r}) = \left(I_J^0 + I_J^P \cos(\mathbf{Q}_{SM} \cdot \mathbf{r})\right)\left(R_N^0 + R_N^P \cos(\mathbf{Q}_{SM} \cdot \mathbf{r} + \varphi)\right)$$
$$= I_J^0 R_N^0 + R_N^0 I_J^P \cos(\mathbf{Q}_{SM} \cdot \mathbf{r}) + I_J^0 R_N^P \cos(\mathbf{Q}_{SM} \cdot \mathbf{r} + \varphi) + \cdots \quad \text{M5.1}$$

where we allow for an arbitrary phase difference $\varphi$ between modulation in $I_J$ and $R_N$ at $\mathbf{Q}_{SM}$. For any value of $\varphi$, one may neglect the third term if

$$I_J^0 R_N^P \ll R_N^0 I_J^P \iff \frac{R_N^P}{R_N^0} \Big/ \frac{I_J^P}{I_J^0} \ll 1 \quad \text{M5.2}$$

In the experiments presented in the main text we measure the quantity

$$I_c(\mathbf{r}) \propto \left(I_J^0\right)^2 + 2I_J^0 I_J^P \cos(\mathbf{Q}_{SM} \cdot \mathbf{r}) + \cdots \quad \text{M5.3}$$

as shown in Extended Data Fig. 9a. In Extended Data Fig. 9b we show $R_N(\mathbf{r})$ derived as above. By comparing the amplitude of $\tilde{I}_c(\mathbf{q})$ and $\tilde{R}_N(\mathbf{q})$ at $\mathbf{q}=\mathbf{Q}_{SM}$ normalized to their $\mathbf{q}=0$ value one may directly determine $\frac{R_N^P}{R_N^0} \Big/ \frac{I_J^P}{I_J^0}$ as in Extended Data Fig. 9c; we find that this ratio is far less than 1 and hence modulations in $\Delta(\mathbf{r})$ at $\mathbf{Q}_{SM}$ are faithfully represented by those measured in $I_c(\mathbf{r})$.

Inset to Extended Data Fig. 9c is $|\tilde{R}_N(\mathbf{q})|$ showing directly that variations in the $\tilde{R}_N(\mathbf{q})$ near $\mathbf{q} \approx (0.25, 0)2\pi/a_0$ and equivalent are negligible so that measurement of $I_c(\mathbf{r})$ should and does yield a faithful image of the condensate.

## (VI) Imaging d-symmetry CDW $\tilde{D}(\mathbf{q})$ and Avoiding CDW Setup Effect in $I_c(\mathbf{r})$

The *d*-symmetry form factor density wave (dFF-DW) in underdoped cuprates (Ref. 39-41) has been directly detected by SI-STM by analyzing the tunneling current $I(\mathbf{r},V)$ and the differential conductance $dI/dV \equiv g(\mathbf{r},V)$ or their ratio maps such as $R(\mathbf{r},V) \equiv I(\mathbf{r},+V)/I(\mathbf{r},-V)$ and $Z(\mathbf{r},V) \equiv g(\mathbf{r},V)/g(\mathbf{r},-V)$. We evaluate $\tilde{D}(\mathbf{q}) = \tilde{O}_x(\mathbf{q}) - \tilde{O}_y(\mathbf{q})$, the function which directly extracts the dFF-DW signature at the correct wavevector from any unprocessed data image that contains sufficient sub-unit-cell resolution. At dopings near optimal, it is known from several experimental techniques that the intensity of the dFF-DW diminishes rapidly. Therefore, to improve the visualization of the dFF-DW signature at such high dopings as used here, we make use of topographic information instead. In constant-current *topographic imaging*, the STM feedback system adjusts the tip sample separation, $T$, as it scans over the sample surface to maintain a set point current, $I_S$, at a constant applied tip-sample bias $V_S$. The topographic image $T(\mathbf{r})$ can be understood by starting with the equation for the tunneling current,

$$I(\mathbf{r},T,V) = f(\mathbf{r},T) \int_0^{eV} LDOS(\mathbf{r},\epsilon)d\epsilon \qquad \text{M6.1}$$

and assuming that the function $f(\mathbf{r},T)$, which represents the effect of corrugation, work function, and tunneling matrix elements, takes the form

$$f(\mathbf{r},T) = \exp(-\kappa T) \cdot A(\mathbf{r}) \qquad \text{M6.2}$$

where $\kappa$ is the WKB factor and depends on the work functions of the sample and tip. The recorded value of the relative tip-sample separation then takes the form

$$T(\mathbf{r}) = \frac{1}{\kappa}\ln\left[\int_0^{eV_S} LDOS(\mathbf{r},\epsilon)d\epsilon\right] + \frac{1}{\kappa}\ln\left[\frac{I_S}{A(\mathbf{r})}\right]. \qquad \text{M6.3}$$

The essential point is that a high signal-to-noise topographic image obtained by constant-current STM imaging reveals contributions from both the surface structure and variations in the $LDOS(\mathbf{r},E)$ (obviously provided $E < eV_s$). In particular, through this effect a CDW imprints its signature logarithmically onto a topographic image far beyond what would be expected from imperceptible deformations of the lattice.

While a single cuprate topographic image will contain the signature of the dFF-DW, subtracting one topographic image from another taken at opposite bias polarity can further enhance the cuprate CDW signal. This is because the cuprate dFF-DW modulates in anti-phase[41] between empty and filled states and hence a subtraction of topographic images measured near the pseudogap energy ($|V_s|\sim 100$ meV in Eqn. M6.3) formed with biases $+V_s$ and $-V_s$ should amplify the CDW contrast.

In Extended Data Fig. 10a,b we demonstrate this procedure using two high resolution, high signal-noise ratio, topographs $T(\mathbf{r})$ taken at opposite bias polarities (100 mV and -100 mV) with the same conventional W tip; they are spatially registered to each other with picometer precision[40]. Extended Data Fig. 10c is the difference of these topographs, from which the two oxygen sublattice sampled images $O_x(\mathbf{r}); O_y(\mathbf{r})$ are then extracted, and $\widetilde{D}(\mathbf{q}) = \widetilde{O}_x(\mathbf{q}) - \widetilde{O}_y(\mathbf{q})$ calculated. An image of $\widetilde{D}(\mathbf{q})$ reveals a primarily $d$-symmetry form factor density wave because of the existence of four maxima at the CDW wavevector $\mathbf{Q}_C = (\pm 0.22, 0)2\pi/a_0; (0, \pm 0.22)2\pi/a_0$ as in Fig. 4d (Ref. 39,41).

One important implication of the fact that the topographic/CDW modulations have $d$-symmetry form factor is for avoidance of the setup effect when measuring $I_c(\mathbf{r})$. Equation M6.3 shows that the topographic contour $T(\mathbf{r})$ is impacted logarithmically by the term $\int_0^{eV_s} LDOS(\mathbf{r},\epsilon)d\epsilon$. This means the tip-surface distance will be modulated at the same $\mathbf{Q}$ as modulations in $LDOS(\mathbf{r},\epsilon)$. One concern might be that because the CDW modulations produce $T(\mathbf{r})$ (i.e. tip-sample distance) modulations at this $\mathbf{Q}$, this will generate a spurious $I_c(\mathbf{r})$ modulation at the same $\mathbf{Q}$. This would be the infamous 'setup effect' but now impacting SJTM.

However as explained above, because the CDW in cuprates has a strongly predominant $d$-symmetry form factor, any conventional image $T(\mathbf{r})$ and its Fourier transform $\widetilde{T}(\mathbf{q})$, that are formed by adding the sublattice images $O_x(\mathbf{r}); O_y(\mathbf{r}); Cu(\mathbf{r})$ contributions in the conventional form $\widetilde{O}_x(\mathbf{q}) + \widetilde{O}_y(\mathbf{q}) + \widetilde{Cu}(\mathbf{q})$, exhibit the actual modulations at wavevector $\mathbf{Q} = \mathbf{Q}_{BRAGG} \pm \mathbf{Q}_C = (1 \pm 0.22, 0)2\pi/a_0; (0,1 \pm 0.22)2\pi/a_0$. These only become detectable at the correct CDW wavevector $\mathbf{Q}_C$ when one uses a measure of $d$-symmetry form factor modulations: $\widetilde{D}(\mathbf{q}) = \widetilde{O}_x(\mathbf{q}) - \widetilde{O}_y(\mathbf{q})$. It is for this reason that CDW modulations in the unprocessed topograph $T(\mathbf{r}); \widetilde{T}(\mathbf{q})$ occur at $\mathbf{Q} = (0.78, 0)2\pi/a_0; (0, 0.78)2\pi/a_0$ (dashed red circles in the Extended Data Fig. 11a). Because they are at a completely different wavevector, it is impossible for such modulations to produce, through a setup effect, spurious $I_c(\mathbf{r})$ modulations at $\mathbf{Q} = (0.25, 0)2\pi/a_0; (0, 0.25)2\pi/a_0$. Thus the PDW wavevector $\mathbf{Q}_P$ observed directly by SJTM, and shown in the Extended Data Fig. 11b and in Fig. 4b, is not generated spuriously by a systematic setup effect due to the coexisting CDW.

## (VII) Analyzing PDW within Ginzburg Landau Theory

Pair density wave (PDW) modulations may be induced by the coupling between a translationally invariant superconducting (SC) order parameter and a charge density wave (CDW). To put this in the context of our data, here we focus on the case of a tetragonal system with **q**=0 SC order of $B_{1g}$ ($d_{x^2-y^2}$) symmetry, a CDW order parameter with a purely *d*-symmetry form factor[42] and an induced secondary PDW order parameter with *s/s'* symmetry form factor. We can then expand the pairing amplitude in terms of superconducting order parameters via

$$\phi(\mathbf{r}, \mathbf{r}') \equiv \langle \psi_\sigma(\mathbf{r})\psi_{-\sigma}(\mathbf{r}')\rangle = D(\mathbf{r} - \mathbf{r}')\Delta_0(R)$$
$$+ S'(\mathbf{r} - \mathbf{r}')\{\Delta_\mathbf{Q}(R)e^{i\mathbf{Q}\cdot\mathbf{R}} + \Delta_{-\mathbf{Q}}(R)e^{-i\mathbf{Q}\cdot\mathbf{R}} + \Delta_{\bar{\mathbf{Q}}}(R)e^{i\bar{\mathbf{Q}}\cdot\mathbf{R}} + \Delta_{-\bar{\mathbf{Q}}}(R)e^{-i\bar{\mathbf{Q}}\cdot\mathbf{R}}\}$$

M7.1

Where $R = (\mathbf{r} - \mathbf{r}')/2$, $D(\mathbf{r} - \mathbf{r}')$ and $S'(\mathbf{r} - \mathbf{r}')$ are the *d*- and *s'*-form factors that either do or do not change sign under 90° rotations. This expansion then defines five complex order parameters that enter the free energy: $\Delta_0, \Delta_\mathbf{Q}, \Delta_{-\mathbf{Q}}, \Delta_{\bar{\mathbf{Q}}}$ and $\Delta_{-\bar{\mathbf{Q}}}$. $\Delta_\mathbf{Q}$ and $\Delta_{-\mathbf{Q}}$ are required to be independent to implement the U(1) spatial phase degree of freedom because $\phi(\mathbf{r}, \mathbf{r}')$ itself is complex. $\mathbf{Q}$ and $\bar{\mathbf{Q}}$ are along symmetry equivalent orthogonal directions. Under a 90° rotation the $B_{1g}$ SC and *s/s'* form factor PDW order parameters transform as $\Delta_0 \to -\Delta_0$ and $\Delta_\mathbf{Q} \to \Delta_{\bar{\mathbf{Q}}}, \Delta_{\bar{\mathbf{Q}}} \to \Delta_{-\mathbf{Q}}, \Delta_{-\mathbf{Q}} \to \Delta_{-\bar{\mathbf{Q}}}, \Delta_{-\bar{\mathbf{Q}}} \to \Delta_\mathbf{Q}$.

In a similar way, the charge density wave order parameter is

$$\phi_c(\mathbf{r}, \mathbf{r}') \equiv \langle \psi^\dagger_\sigma(\mathbf{r})\psi_\sigma(\mathbf{r}') + \psi^\dagger_\sigma(\mathbf{r}')\psi_\sigma(\mathbf{r})\rangle$$
$$= D(\mathbf{r} - \mathbf{r}')\{\psi_\mathbf{Q}(R)e^{i\mathbf{Q}\cdot\mathbf{R}} + \psi_{-\mathbf{Q}}(R)e^{-i\mathbf{Q}\cdot\mathbf{R}} + \psi_{\bar{\mathbf{Q}}}(R)e^{i\bar{\mathbf{Q}}\cdot\mathbf{R}}$$
$$+ \psi_{-\bar{\mathbf{Q}}}(R)e^{-i\bar{\mathbf{Q}}\cdot\mathbf{R}}\}\{\psi_\mathbf{Q}(R)e^{i\mathbf{Q}\cdot\mathbf{R}} + \psi_{-\mathbf{Q}}(R)e^{-i\mathbf{Q}\cdot\mathbf{R}} + \psi_{\bar{\mathbf{Q}}}(R)e^{i\bar{\mathbf{Q}}\cdot\mathbf{R}}$$
$$+ \psi_{-\bar{\mathbf{Q}}}(R)e^{-i\bar{\mathbf{Q}}\cdot\mathbf{R}}\}$$

M.7.2

This will contribute two independent complex order parameters to the free energy expansion. Since the CDW order parameter must be real $\psi_{-\mathbf{Q}} = \psi^*_\mathbf{Q}$. For purely *d*- symmetry form factor CDW, the order parameter will have the following property under a 90° rotation : $\psi_Q \rightarrow -\psi_{\bar{Q}}, \psi_{\bar{Q}} \rightarrow -\psi_{-Q}, \psi_{-Q} \rightarrow -\psi_{-\bar{Q}}, \psi_{-\bar{Q}} \rightarrow -\psi_Q$. Now consider the symmetry-allowed free energy terms that will couple $\Delta_0$, the $\Delta_Q$s and the $\psi_Q$s. The leading term is simply:

$$\gamma_1 \left[ \Delta_0^* \{ \psi_Q \Delta_{-Q} + \psi_Q^* \Delta_Q + \psi_{\bar{Q}} \Delta_{-\bar{Q}} + \psi_{\bar{Q}}^* \Delta_{\bar{Q}} \} + c.c \right] \qquad \text{M.7.3}$$

Because the PDW order parameters appear to the power of unity in this term, the free energy can always be lowered by making $\Delta_Q$ non-zero, if both $\Delta_0^*$ and $\psi_{\bar{Q}}$ are non-zero. Thus the co-existence of $d_{x^2-y^2}$ superconductivity and a *d*-symmetry form factor CDW at wave-vector **Q** will induce a PDW at wave-vector **Q** with an *s/s'* symmetry form factor.

We note that such a coupling is not unique to this particular set of form factors. There exists a similar symmetry-allowed term in the free energy admitting a *d*-symmetry form factor PDW where the $\Delta_Q$s now transform under 90 degree rotation like the $\psi_Q$s. It is

$$\gamma_2 \left[ \Delta_0^* \{ \psi_Q \Delta_{-Q} + \psi_Q^* \Delta_Q - \psi_{\bar{Q}} \Delta_{-\bar{Q}} - \psi_{\bar{Q}}^* \Delta_{\bar{Q}} \} + c.c \right] \qquad \text{M.7.4}$$

As the experimentally detected PDW has *s/s'* form factor we do not discuss this further here. But its existence demonstrates that the coexistence of all three orders, the uniform SC, CDW and PDW, where CDW and PDW have the same wave vectors due to the presence of a uniform SC, is form factor agnostic in agreement with Refs. 14, 20, 43 .

**(VIII) Data and Code**

The data files for the results presented here and the codes used in their analysis are available at http://dx.doi.org/10.17630/83957a84-2186-4c14-8e55-f961a19ec9a9 .

# Extended Data Figure Captions

**Extended Data Figure 1 SJTM Circuit Model and Phase Diffusion Dynamics**
a. Circuit diagram of the hybrid SISTM/SJTM setup used in these experiments. The voltage source $V$ is from the usual STM bias controller; the load resistor $R_B$=10MΩ; the single-particle tunneling resistance of the Josephson junction formed by tip/sample is $R_N$, and the voltage actually developed across the junction is $V_{JJ}$.
b. The dynamics of this circuit produces two predominant effects. First there is a sudden change in the $I(V)$ characteristic measured with the external ammeter shown, when the current reaches a value $I_c = \frac{\hbar}{8ekT^*}I_J^2$ where $I_J$ is the zero-temperature Josephson critical current and $T^*$ an effective temperature parameterizing the dissipative environment of the junction. The second predicted effect is strong hysteresis depending on which direction the external voltage is swept; this is shown as the difference between solid-red and dashed-blue lines. Both effects are seen very clearly and universally in the measured $I(V)$ throughout our studies reported here.
c. The relationship of the current as in b but in terms of the voltage across the Josephson junction $V_{JJ}$.

**Extended Data Figure 2 Simulation of Expected SIS d$I$/d$V$($V$) for BSCCO Nano-flake Tip**
Dashed curve represents a typical d$I$/d$V$ spectrum on as-grown $Bi_2Sr_2CaCu_2O_{8+x}$ (BSCCO) sample, measured using a normal metallic W tip. The solid line is a simulation for an expected d$I$/d$V$ spectrum when using a BSCCO superconducting tip; we use the standard equation for tunneling between two superconductors, each with $DOS(E)$ spectrum identical to the dashed line. The result, in very good agreement with the typical measured SIS spectrum, is shown in Fig. 2b.

**Extended Data Figure 3 Spectroscopic/Topographic Data from Two Distinct BSCCO Tips**
Typical d$I$/d$V$ spectrum, topography $T(r)$ and magnitude of its Fourier transform $\tilde{T}(q)$ measured with two completely distinct BSCCO tips on two different BSCCO samples: tip 1, a c e; tip 2 b d f.

**Extended Data Figure 4 Simulation of Topography with BSCCO Nano-flake Tip**
a. Surface topography $T(r)$ of BSCCO sample; image obtained with a conventional metallic W tip.
b. Magnitude of Fourier transform of a.
c. Model for BSCCO nano-flake tip that is aligned with the BSCCO surface in a.
d. Model for BSCCO nano-flake tip that is misaligned with the BSCCO surface in a.
e. Convolution of BSCCO surface image in a and aligned BSCCO nano-flake tip in c. Inset shows resultant Fourier Transform with Bragg peaks in the corners and supermodulation peaks.
f. Convolution of BSCCO surface image in a and misaligned BSCCO nano-flake tip in d. Inset shows resulting Fourier transform with many additional broad peaks caused by Moiré pattern effects. Comparison to e demonstrates that the tip used in the studies reported here was well aligned with sample.

**Extended Data Figure 5 Spatial Variation of HTS-Tip/Sample Josephson Junction *I*(*V*)**
a. A histogram of all $I_c$ values measured at different locations in the field of view shown in Fig 3b. The characteristic $I_c$ values associated with the three spectra in b are indicated by the colored arrows.
b. Three *I*(*V*) formed by averaging the constituent *I*(*V*) from the bins indicated by the colored arrows in a. They demonstrate the variation in measured *I*(*V*) characteristic that occur at different locations in the field of view shown in Fig 3b. The three $I_c$ values are indicated by colored arrows.

**Extended Data Figure 6 Repeatable Utility of BSCCO Nano-flake Tips for *I*$_c$(*r*) Mapping**
Three *I*$_c$(*r*) images a,b,c measured with same BSCCO nano-flake tip, at different times (separated by many days) and using different Josephson junction parameters, but in closely related in fields-of-view. Repeatability and fidelity of our *I*$_c$(*r*) imaging by SJTM is evident.

**Extended Data Figure 7 Sequence of measured *I*$_c$(*r*) along line in Extended Data Fig. 6b**

*28*

Nine individual I(V) spectra measured at the locations indicated by numbers 1-9 in Extended Data Fig. 6b . In each case the transition from the SIS resistive branch to the phase diffusion Josephson tunneling branch is evident. Moreover as the sequence passes through the site of a Zn atom, the value of $I_c$ diminishes by ~95 % from its maximum, as expected from muon spin rotation experiments. .

**Extended Data Figure 8 Before/After Topographic images Bracketing $I_c$(r) Map**
a. Topograph taken with BSCCO tip before typical $I_c$(**r**) SJTM map.
b. Topograph taken with same BSCCO tip after the same $I_c$(**r**) SJTM map as a. Comparison of a,b shows that tip and surface are very well preserved in our SJTM protocol.

**Extended Data Figure 9 Comparison of Modulations in $I_c$(r) and $R_N$(r)**
a. A typical measured $I_c(\bm{r})$ image of Bi$_2$Sr$_2$CaCu$_2$O$_8$ with the crystal supermodulation effect retained and apparent as strong spatial modulations in $I_c$ along the vertical axis.
b. Measured $R_N(\bm{r})$ image simultaneous with a, with the crystal supermodulation effect retained. The spatial modulations in $R_N(\bm{r})$ along the vertical axis are greatly diminished in relative amplitude compared to $I_c(\bm{r})$ modulations in a.
c. Inset shows $|\tilde{R}_N(\bm{q})|$ the magnitude of the Fourier transform of $R_N(\bm{r})$ from b. Plotting the simultaneously measured Fourier amplitudes of $\tilde{I}_c(\bm{q})$ and $\tilde{R}_N(\bm{q})$ along the (1,1) direction passing through the wavevector of the supermodulation **Q**$_{SM}$, shows that modulations in $\tilde{R}_N(\bm{q})$ are negligible. Therefore the predominant effect in the $I_c(\bm{r})R_N(\bm{r})$ studied through this work is due to the $I_c(\bm{r})$ variations coming from visualizing the superfluid density variations $\rho_S(r) \propto I_J^2(r) \propto I_c(r)$ of the condensate in the sample.

**Extended Data Figure 10 d-symmetry Density Wave $\tilde{D}(q)$ from Topography**
a. High resolution topographic image of typical BiO surface at the same hole-density as the $I_c(\bm{r})$ studies, measured at V=100 meV
b. High resolution topographic image of identical (registered for every atom within about 10 pm) BiO surface as a measured at V=-100 meV

c. Difference between a and b; a CDW exhibits its signature logarithmically in such an image and therefore it can be used to detect the d-symmetry form factor density wave as in Fig. 4d of the main text.

**Extended Data Figure 11 Absence Charge Density Wave Setup Effect in $I_c(r)$**

a. $\tilde{T}(q)$, the Fourier transform magnitude of the sublattice-resolved image $O_x(r) + O_y(r)$ derived from $\delta T(r)$ the differences between the two unprocessed topographic images $T(r,\pm100$ meV$)$ in Extended Data Fig 10. We see directly that the actual modulations in topography due to the density of states modulations from the CDW, occur at wavevectors $(1 \pm 0.22, 0)2\pi/a_0; (0,1 \pm 0.22)2\pi/a_0$ (dashed circles), as has been reported extensively in the past. These only become detectable at the actual CDW wavevector $Q_C = (0.22,0)2\pi/a_0; (0,0.22)2\pi/a_0$ when one uses a measure of d-symmetry form factor modulations: $\tilde{D}(q) = \tilde{O}_x(q) - \tilde{O}_y(q)$ as shown in Fig. 4d of main text. Because the physically real modulations in topography and conductance imaging therefore occur at $Q = (0.78,0)2\pi/a_0; (0,0.78)2\pi/a_0$ (dashed circles), it is impossible for them to produce, through a 'setup effect', spurious $I_c(r)$ modulations at the PDW wavevector $Q_P = (0.25,0)2\pi/a_0; (0,0.25)2\pi/a_0$ as indicated by dashed circles in b.

b. The measured q-space structure $\tilde{I}_c(q)$ (which samples all sublattices in the conventional form $\tilde{O}_x(q) + \tilde{O}_y(q) + \tilde{C}u(q)$. The PDW maxima occur at $Q_P = (0.25,0)2\pi/a_0; (0,0.25)2\pi/a_0$.

EXT. DATA FIG 1

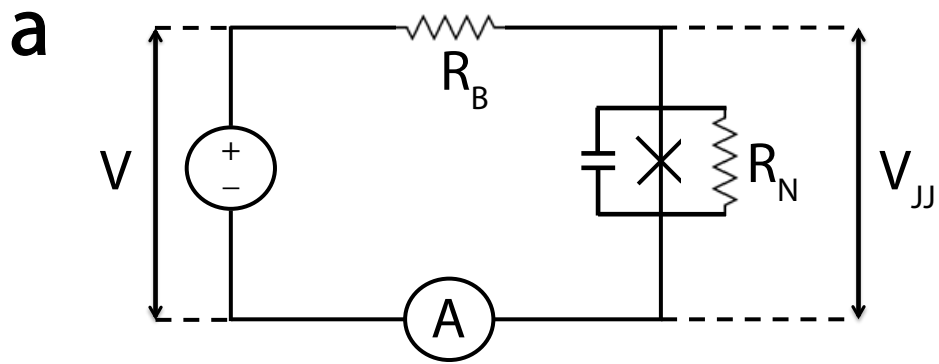
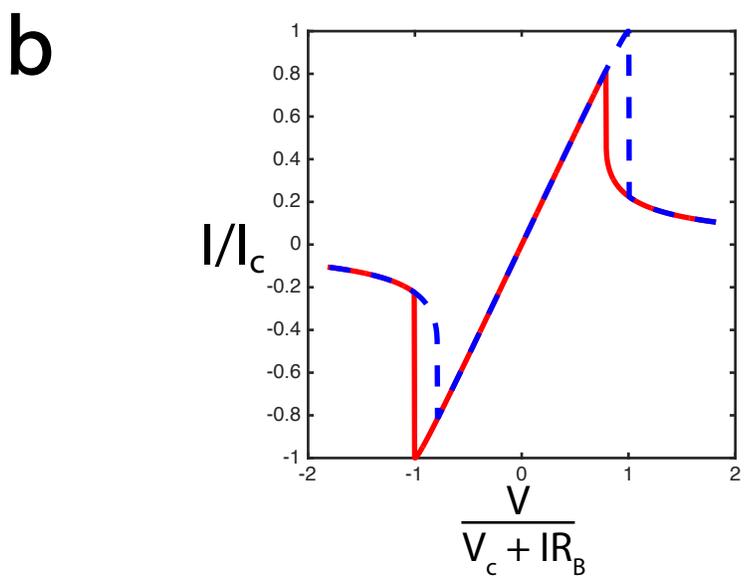
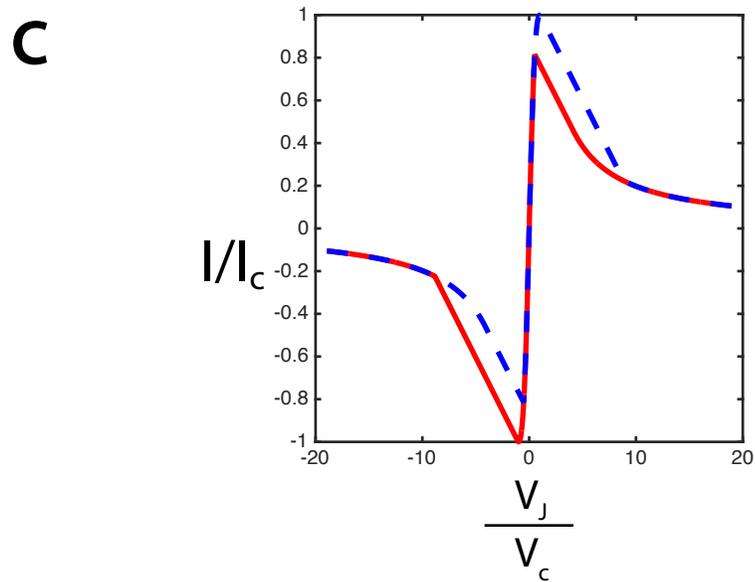

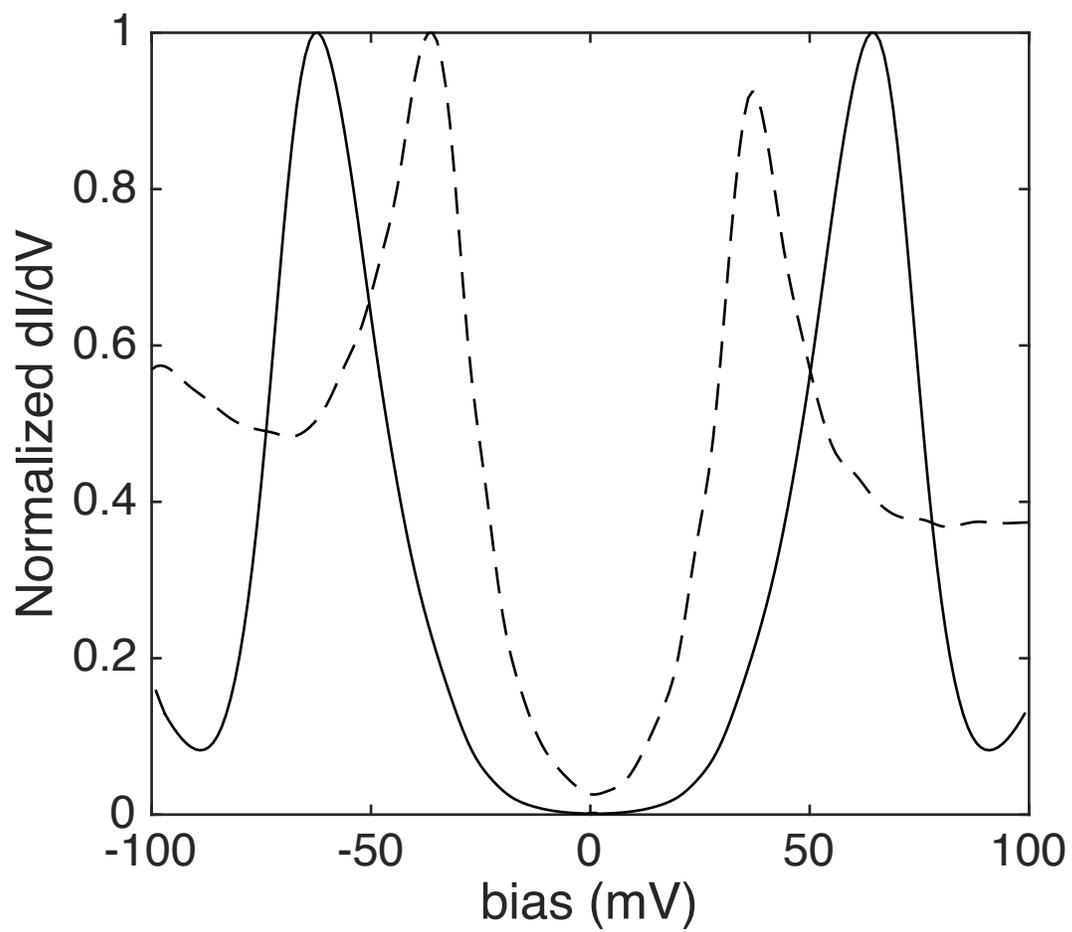

EXT. DATA FIG 3

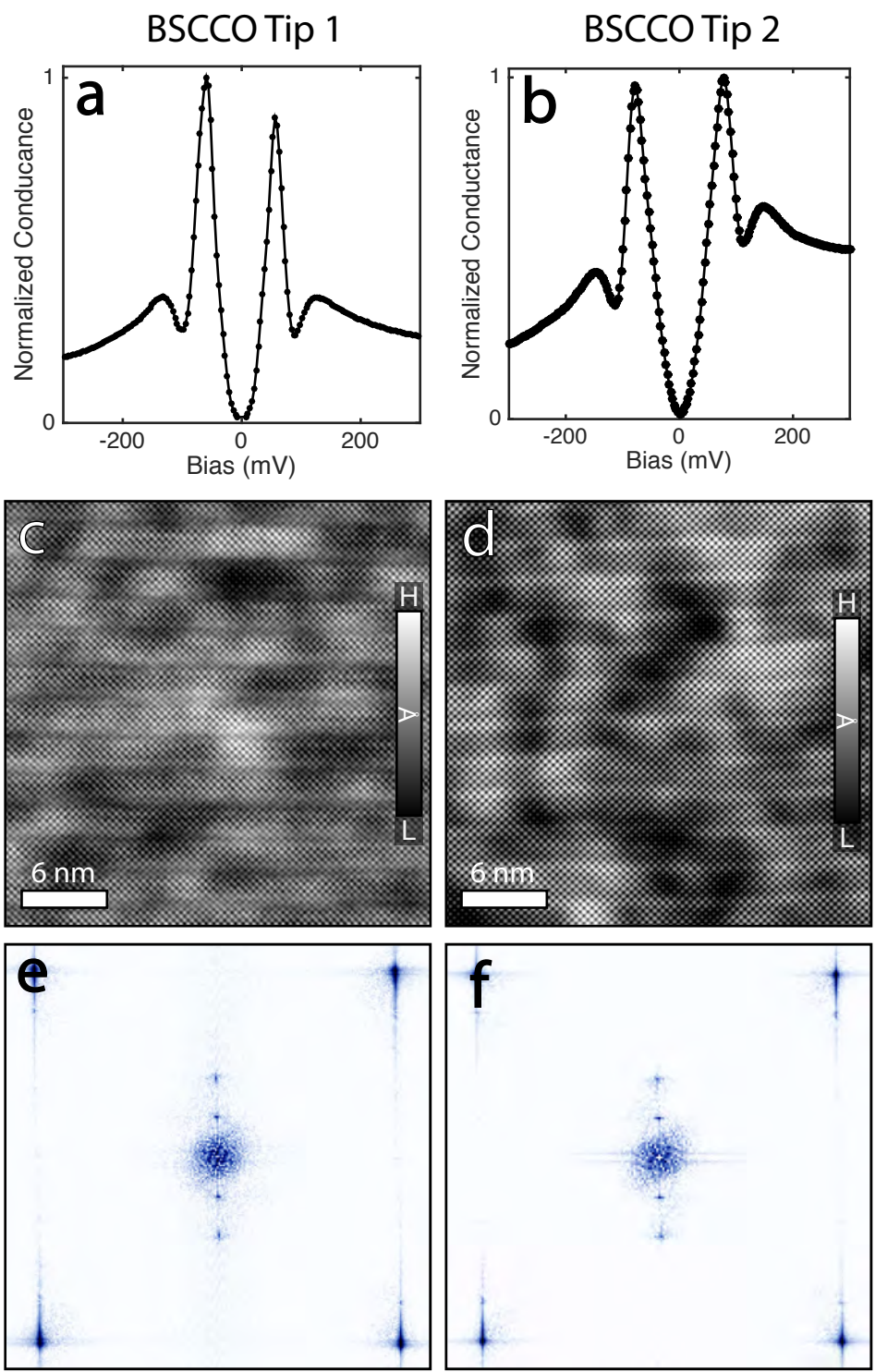

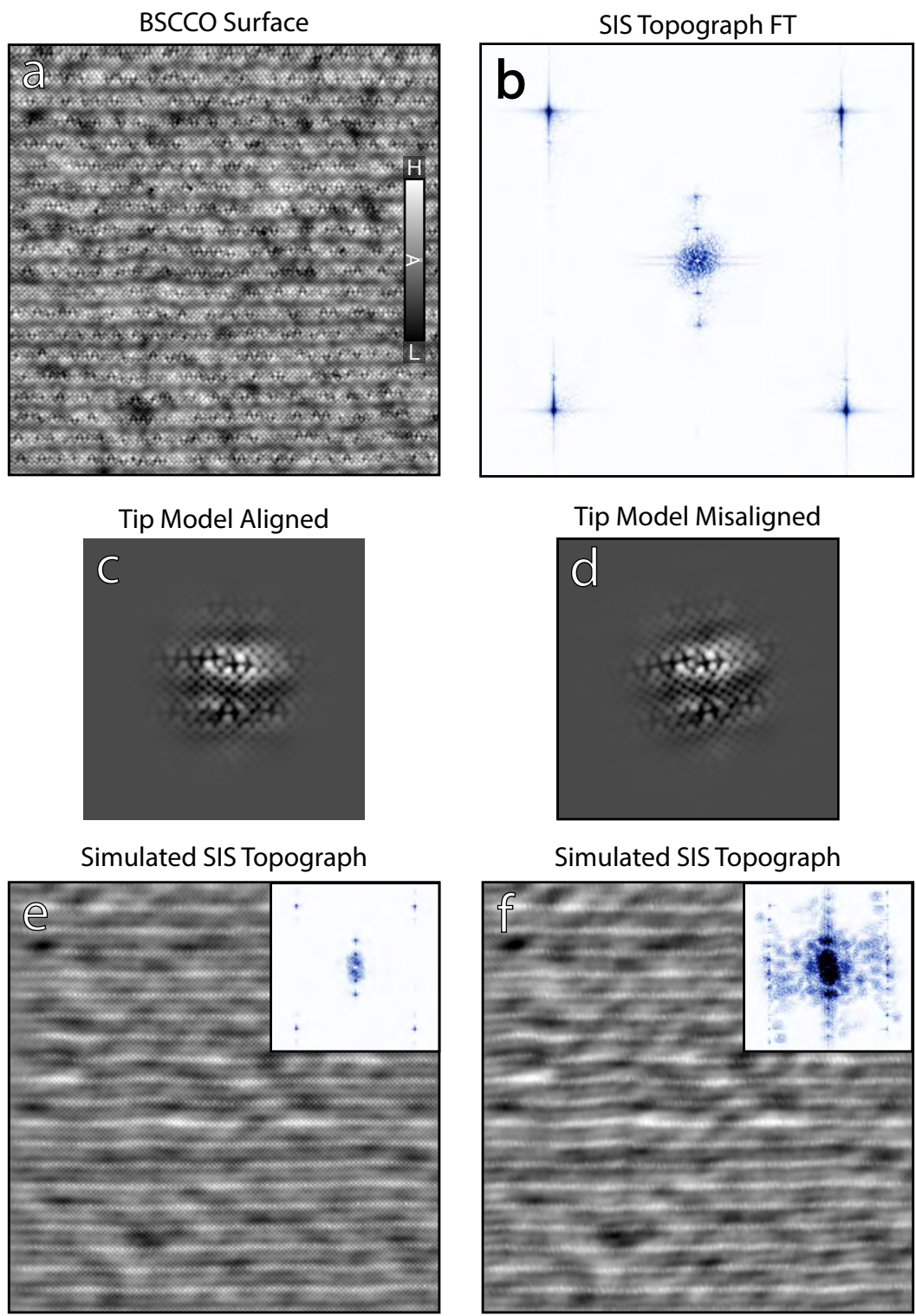

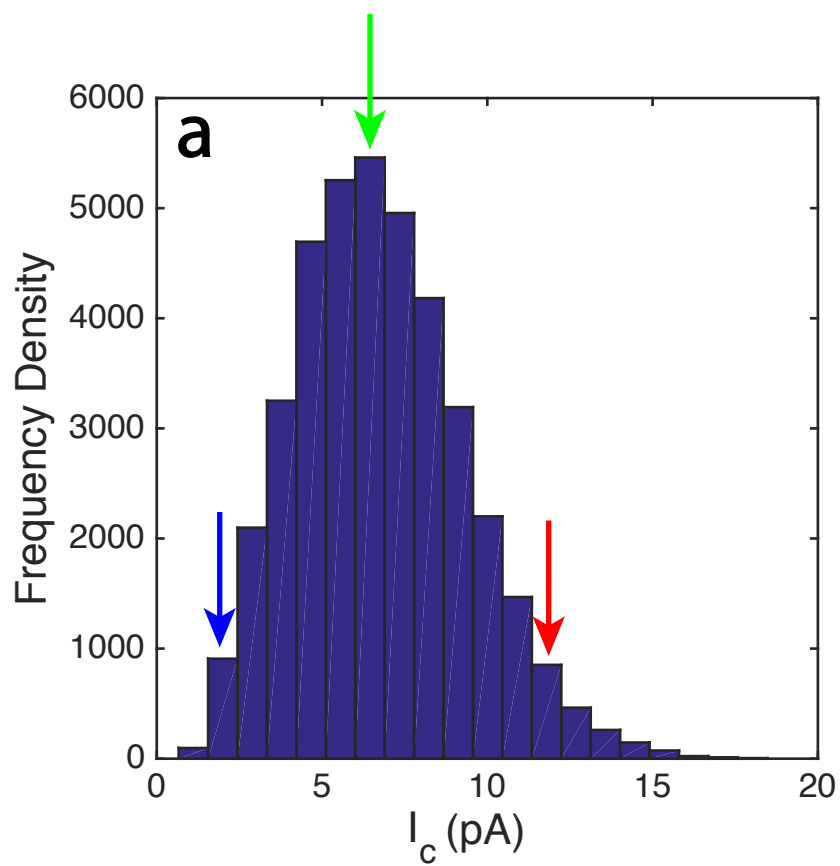

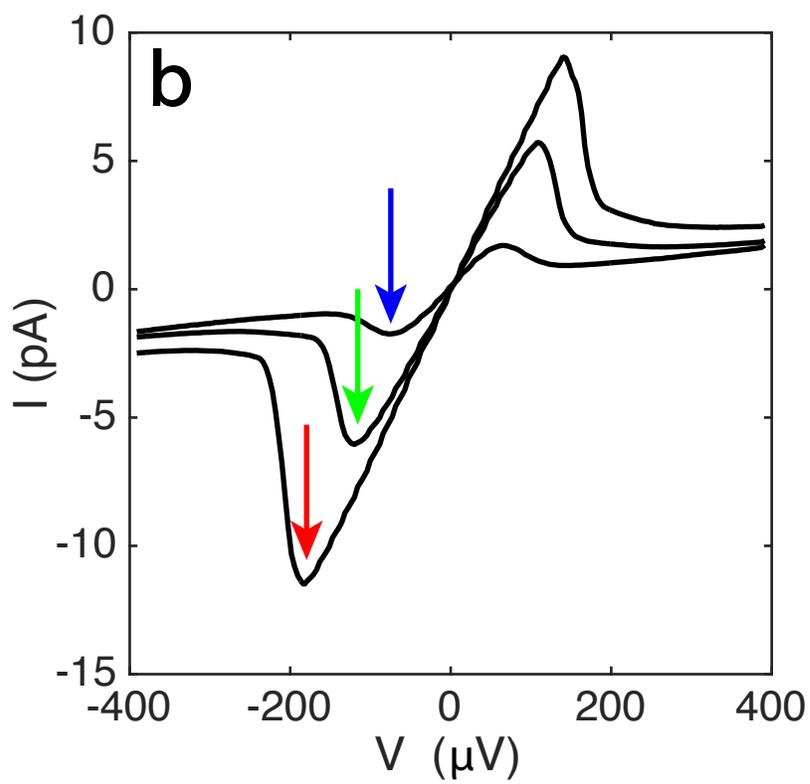

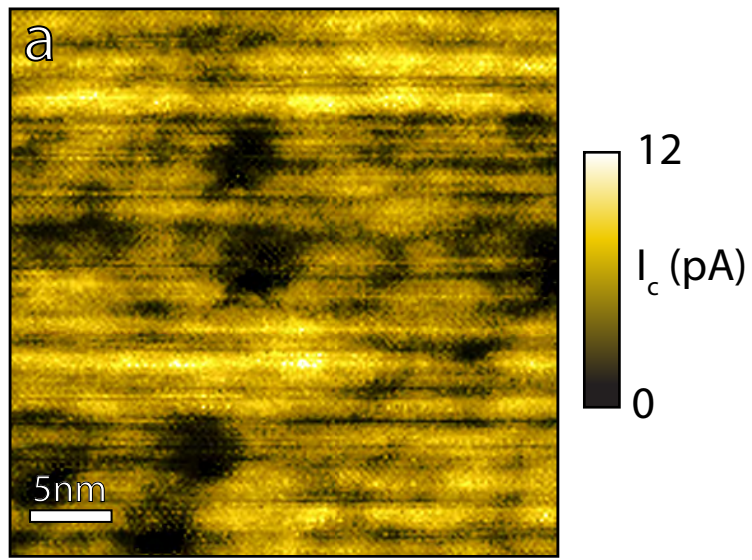
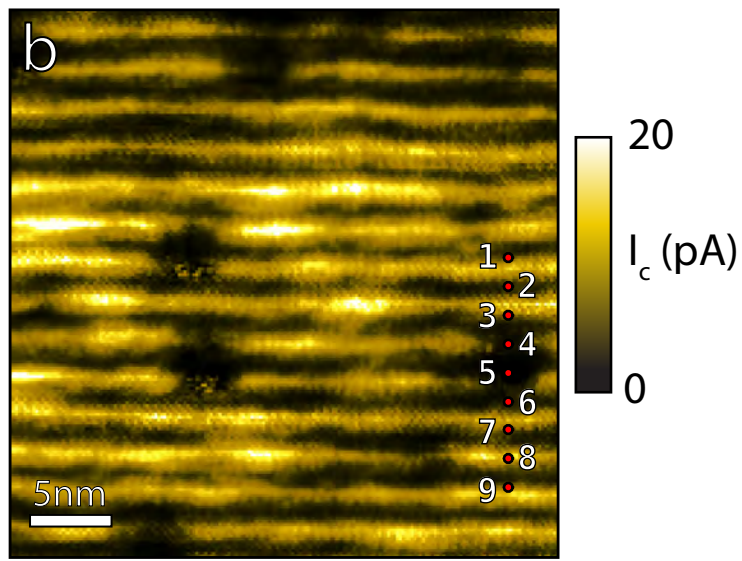
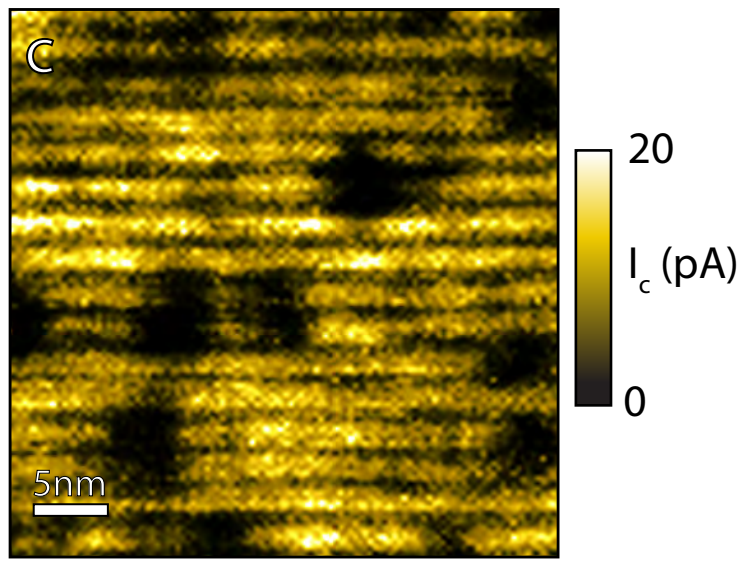

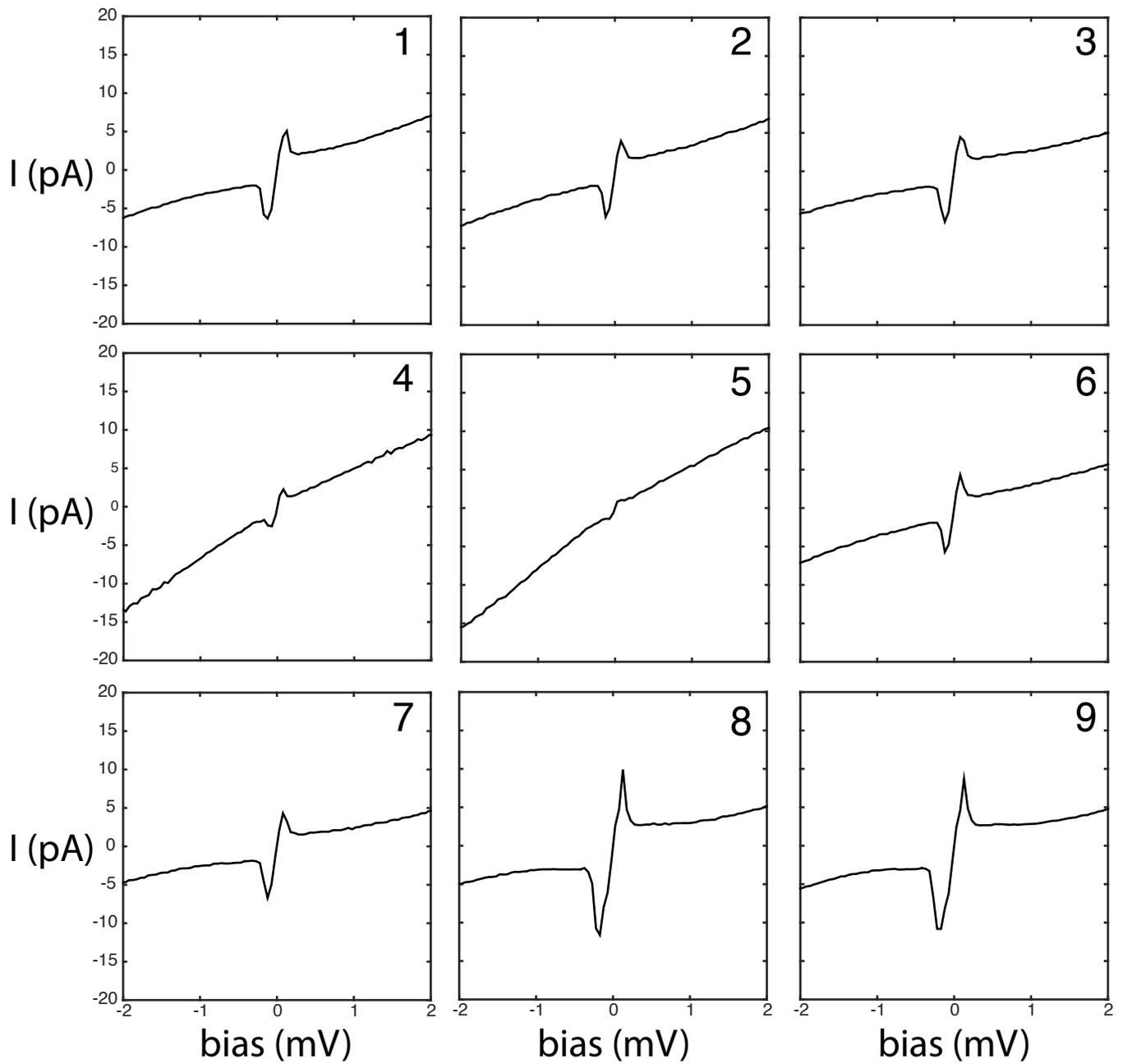

EXT. DATA FIG 8

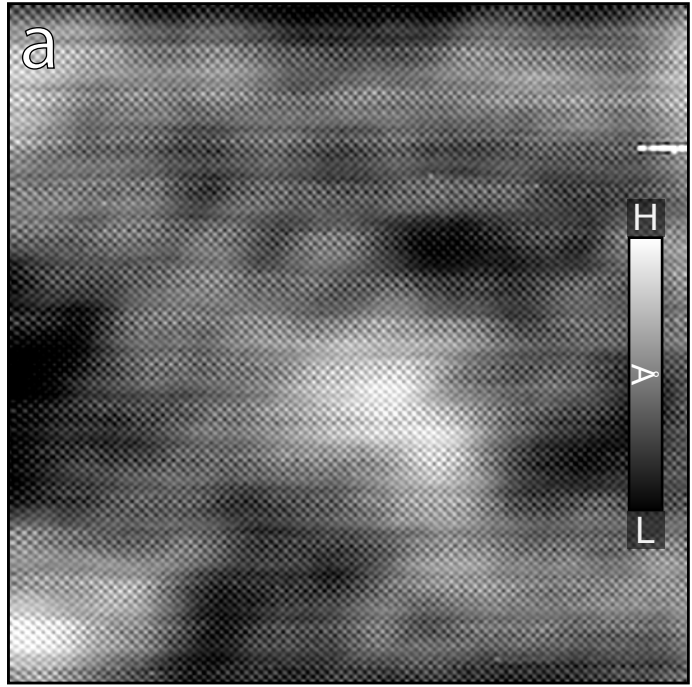

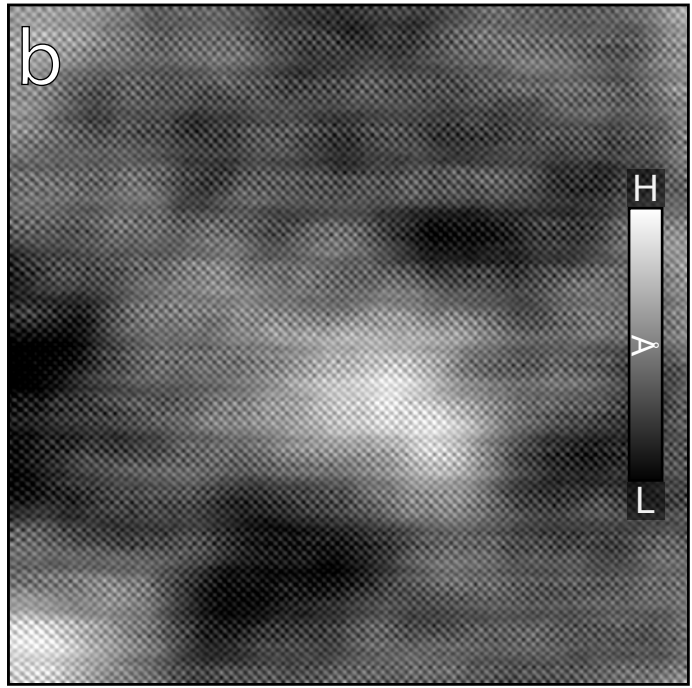

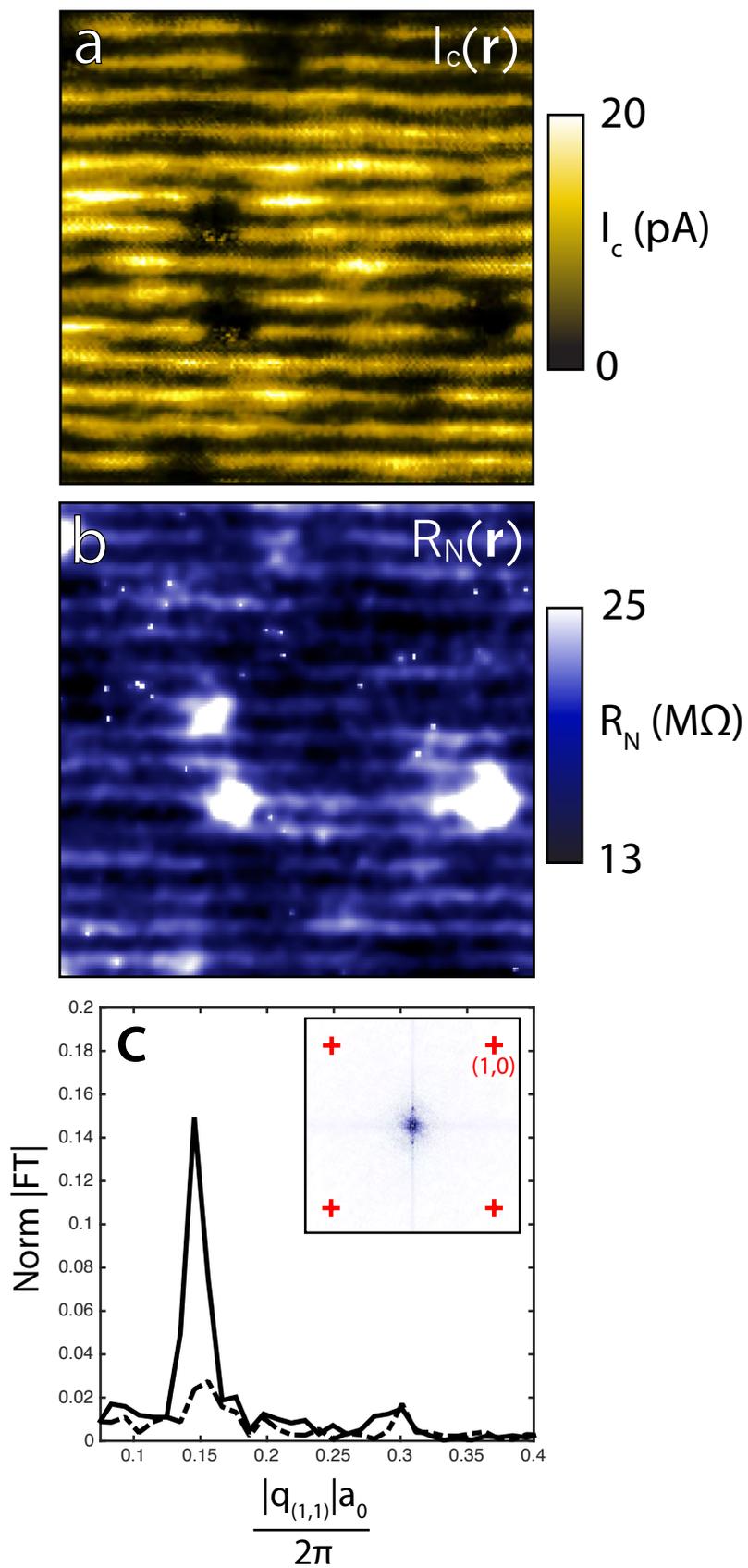

EXT. DATA FIG 10

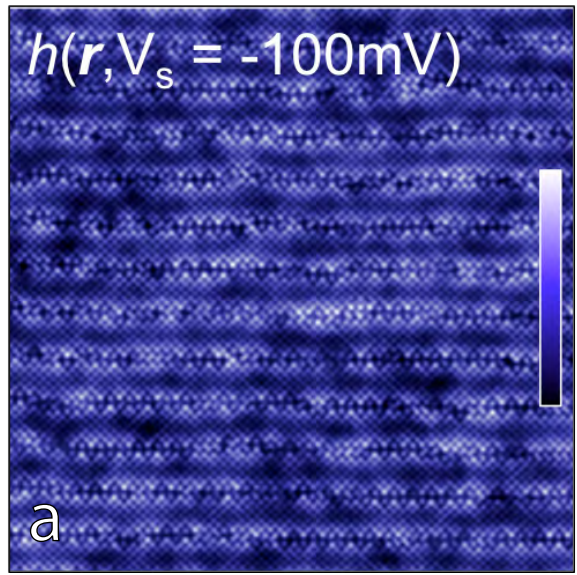

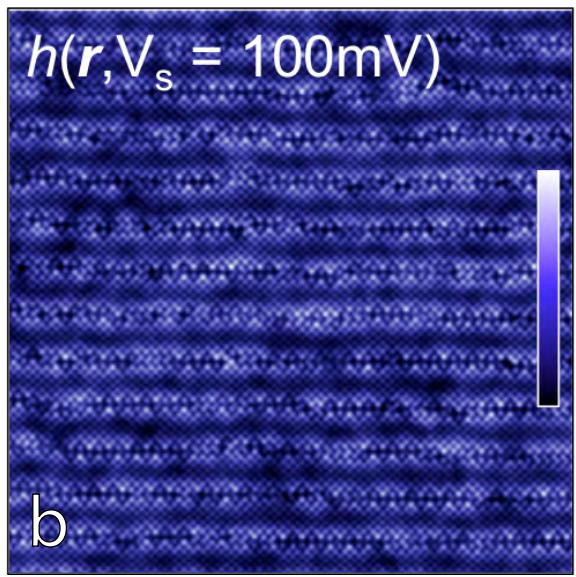

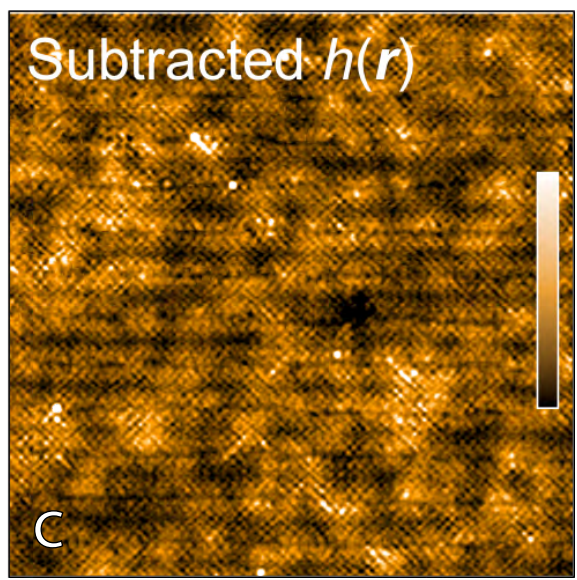

EXT. DATA FIG 11

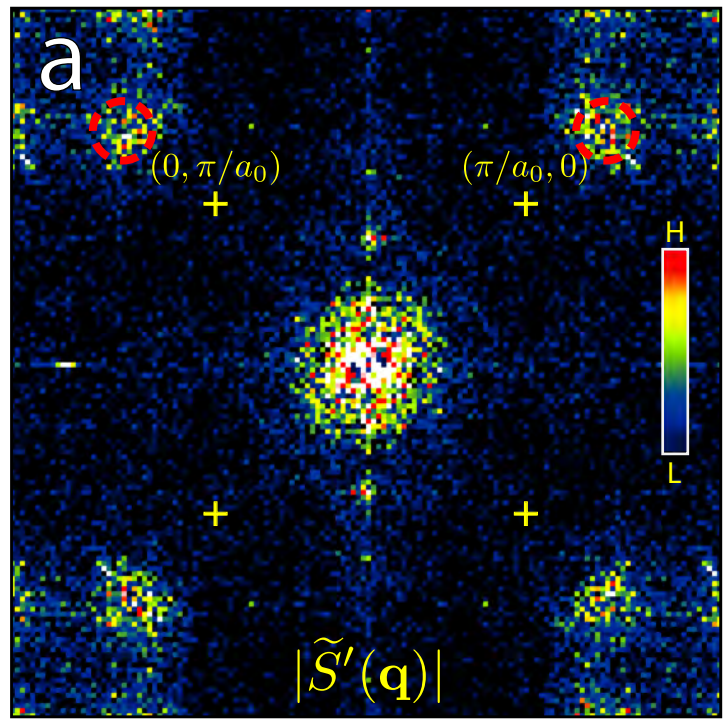

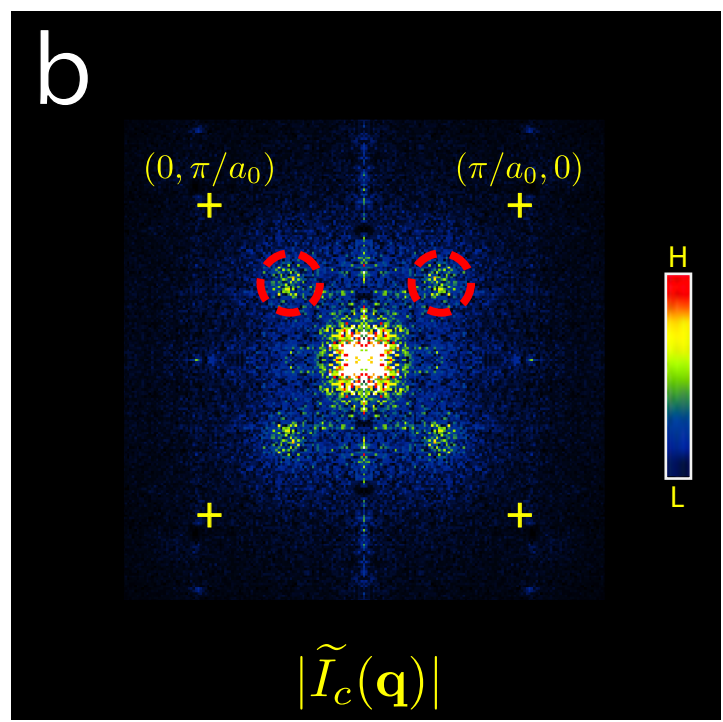